\documentclass[aps,prd,twocolumn,superscriptaddress,nofootinbib]{revtex4}
\usepackage{amsmath}
\usepackage{siunitx}
\usepackage{enumitem}
\usepackage{multirow}

\usepackage{graphicx}
\usepackage[section]{placeins}
\usepackage[usenames]{xcolor}

\usepackage{hyperref}

\newcommand{\Su}{{\(S_\uparrow\)}}
\newcommand{\So}{{\(S_0\)}}
\newcommand{\SuM}{{\(S^\mathrm{Mag}_\uparrow\)}}
\newcommand{\SoM}{{\(S^\mathrm{Mag}_0\)}}

\begin{document}

\title{Effect of magnetic fields on the dynamics and gravitational wave emission of PPI-saturated self-gravitating accretion disks: simulations in full GR}

\author{Erik Wessel}
\affiliation{Department of Physics, University of Arizona, Tucson, AZ}

\author{Vasileios Paschalidis}
\affiliation{Departments of Astronomy and Physics, University of Arizona, Tucson, AZ}

\author{Antonios Tsokaros}
\affiliation{Department of Physics, University of Illinois, Urbana-Champaign, IL}
\affiliation{National Center for Supercomputing Applications, University of Illinois at Urbana-Champaign, Urbana, IL}
\author{Milton Ruiz}
\affiliation{Departamento de Astronom\'{\i}a y Astrof\'{\i}sica, Universitat de Val\`encia,
  Dr. Moliner 50, 46100, Burjassot (Val\`encia), Spain}
\author{Stuart L. Shapiro}
\affiliation{Departments of Physics and Astronomy, University of Illinois, Urbana-Champaign, IL}

\date{\today}

\begin{abstract}
We explore the effect magnetic fields have on self-gravitating
accretion disks around spinning black holes via numerical evolutions
in full dynamical magnetohydrodynamic spacetimes.  The configurations
we study are unstable to the Papaloizou-Pringle Instability (PPI).
PPI-saturated accretion tori have been shown to produce gravitational
waves, detectable to cosmological distances by third-generation
gravitational wave (GW) observatories.  While the PPI operates
strongly for purely hydrodynamic disks, the situation can be different
for disks hosting initially small magnetic fields.  Evolutions of
disks without self-gravity in fixed BH spacetimes have shown that
small seed fields can initiate the rapid growth of the
magneto-rotational instability (MRI), which then strongly suppresses
the PPI.  Since realistic astrophysical disks are expected to be
magnetized, PPI-generated GW signals may be suppressed as well.
However, it is unclear what happens when the disk self-gravity is
restored. Here, we study the impact of magnetic fields on the
PPI-saturated state of a self-gravitating accretion disk around a
spinning BH ($\chi = 0.7$) aligned with the disk angular momentum, as
well as one around a non-spinning BH.  We find the MRI is effective at
reducing the amplitude of PPI modes and their associated GWs, but the
systems still generate GWs. Estimating the detectability of these
systems accross a wide range of masses, we show that magnetic fields
reduce the maximum detection distance by Cosmic Explorer from 300Mpc
(in the pure hydrodynamic case) to 45Mpc for a $10 M_{\odot}$ system,
by LISA from 11500Mpc to 2700Mpc for a $2 \times 10^{5} M_{\odot}$
system, and by DECIGO from $z \approx 5$ down to $z \approx 2$ for a
$1000 M_{\odot}$ system.
\end{abstract}


\maketitle

\section{Introduction\label{section:intro}}
Thick gaseous tori that orbit and accrete onto black holes have been a
major topic of study in astrophysics for decades.
Early theoretical work on these systems was motivated by the
need to model X-ray binaries and active galactic
nuclei~\cite{rees_black_1984}.  Supermassive star collapse~\cite{shibata_collapse_2002}, tidal
disruption events by black holes, mergers of binary neutron stars, and
of binary black hole-neutron stars~\cite{lovelace_massive_2013,paschalidis_relativistic_2015} can all
form short-lived (massive) accretion tori. Thus black hole (BH) disk
systems are known to be ubiquitous in the universe, and much work has
been done to model the electromagnetic signatures of these objects.
However, with the rise of gravitational wave (GW)
astronomy~\cite{prospects_2020} and the promise of third generation GW
observatories to open up new frequency ranges and achieve greater
sensitivity~\cite{reitze_cosmic_2019,amaro-seoane_laser_2017,sato_status_2017},
interest has arisen in BH-disk systems as potential
GW sources.

In order to produce GW radiation, an astrophysical object must have
significant time variation of its quadrapole moment.\footnote{GWs may
  also be produced by changes in higher-order mass and angular
  momentum moments, but in practice these are typically sub-dominant
  contributions.}  For BH-disk systems, the simplest way for this to
occur is for the disk to be non-axisymmetric, with a significant
concentration of mass in one or more orbiting lumps that persist for
many orbits.  The well known hydrodynamic Papaloizou-Pringle
Instability (PPI)~\cite{papaloizou_dynamical_1984} saturates to 
such a quasi-steady-state configuration~\cite{hawley_non-linear_1987}.  The PPI is
a global instability with modes that grow exponentially in amplitude
from initially small perturbations in axisymmetric disks. The growth
of PPI modes can be understood intuitively as a run-away feedback loop
resulting from the exchange of a conserved--but \textit{not}
lower-bounded--quantity between wave-like disturbances on the inner
edge of the disk which propagate opposite the flow, and wave-like
disturbances on the outer edge of the disk which propagate with the
flow~\cite{blaes_oscillations_1985,blaes_stability_1986,goldreich_stability_1986,narayan_physics_1987,goodman_stability_1988,christodoulou_stability_1992}.
The fastest-growing modes of the PPI are simple non-axisymmetric
perturbations of the rest-mass density $\rho_{0}$ of the form,
\begin{equation}\label{eq:m_modes}
  \rho_{0} \propto e^{i(m\phi - \sigma t)},
\end{equation}
where $\phi$ is the azimuthal angular coordinate and $m$ is the
integer mode number, which determines the number of over-dense lumps
produced in the disk, $\sigma$ is a different complex number for each
$m$, with a real component that controls the pattern's orbital
frequency, and an imaginary component that controls its exponential
growth rate.  The low-$m$ modes are fastest growing, with patterns
that rotate at roughly the orbital frequency of the maximum-density
region of the disk.

Suceptability to the PPI is determined by the specific angular momentum profile of the initially axisymmetric disk.
In a Newtonian context, the specific angular momentum is defined as
\begin{equation}\label{eq:j_newt}
  j \equiv rv_{\phi},
\end{equation}
where $r$ is the cylindrical radius and $v_\phi$ the azimuthal
component of the velocity.  For axisymmetric disks \(j\) typically is
nearly a power-law in terms of \(r\), and the criteria for PPI
instability is \cite{papaloizou_dynamical_1985},
\begin{equation}\label{eq:PPI_criterion}
  j \propto r^{2-q},\sqrt{3} < q \le 2.
\end{equation}
Disks with \(q > 2\) have radially decreasing specific angular
momentum, and are quickly accreted due to a runaway axisymmetric
instability first described by Rayleigh \cite{strutt_dynamics_1917}:
we will not concern ourselves with such disks.  While the criteria
above were originally derived in the Newtonian regime, it turns out
that these criteria can still fruitfully be applied to the PPI in
relativistic disks~\cite{kojima_dynamical_1986}, provided we use a
relativistic definition for the specific angular momentum,
\begin{equation}\label{eq:spec_ang_def}
  j \equiv u^{t}u_{\phi},
\end{equation}
where $u^{\mu}$ is the fluid 4-velocity in a polar coordinate system.

The PPI was first studied for a self-gravitating disk orbiting a
non-spining BH in full, dynamical spacetime numerical relativity
in~\cite{korobkin_stability_2011}.  Shortly thereafter longer evolutions of a similar BH
disk system were published by~\cite{kiuchi_gravitational_2011}, who
were able to extract long GW waveforms produced by the PPI, and
assessed the potential detectability of such signals.
Smoothed-particle hydrodynamics simulations of tori produced by tidal
disruption events demonstrated one astrophysically plausible pathway
for producing PPI-unstable disks~\cite{nealon_papaloizoupringle_2018},
although the estimates of GWs produced by those
disks~\cite{toscani_gravitational_2019} show that the amplitude would
be much lower than the configuration studied
in~\cite{kiuchi_gravitational_2011}. In~\cite{mewes_numerical_2016} PPI-unstable self-gravitating disks were evolved around
tilted, spinning black holes, but starting from constraint-violating initial data.

In~\cite{wessel_gravitational_2021} we studied the PPI in
self-gravitating disks around spinning BHs with constraint-satisfying initial data for the first time.
Three nearly identical BH disk systems were evolved, differing
primarily in the spin of the central BH, which in one case had no spin,
and in the other two had a dimensionless spin magnitude of $\chi = 0.7$ either
aligned or anti-aligned with the disk orbital angular momentum.  We
found that BH spin only altered the PPI dynamics through the changes
induced in disk orbital angular frequency, but that the behavior of
the instability was generally unchanged.  However, in the aligned-spin
case, the reduction in the innermost stable circular orbit (ISCO)
radius slowed accretion and improved signal duration.  We also carried
out a thorough assessment of detectability, showing that such a system
could potentially be detected out to cosmological distances by
next-generation space-based GW
observatories. In~\cite{tsokaros_self-gravitating_2022} we followed up with
hydrodynmic simulations of self-gravitating tori tilted with respect
to the BH spin, where we found that PPI saturates earlier and these
systems generate GWs beyond the dominant (2,2)
mode. In~\cite{shibata_alternative_2021} GWs produced by the growth of
low-$m$ non-axisymmetries in thick tori with shallow $j$ profiles
around spinning BHs were investigated as a plausible alternative
explanation for GW190521.

While these results are exciting from a GW astronomy point of view,
these studies did not account for magnetic fields. Magnetic fields are
important because the magneto-rotational instability (MRI) grows
rapidly in weakly-magnetized accretion
disks~\cite{balbus_powerful_1992}. MRI quickly amplifies small initial
fields and drives turbulence that transports angular momentum which
allows accretion to proceed~\cite{balbus_instability_1998}.

Since the quadrupole moment of the BH-disk system is primarily
dictated by the large-scale bulk motion of the disk, an end-state
configuration produced by the MRI could be unfavorable to detecting
GWs.  Both the PPI and the MRI grow at rates near the orbital
frequency of the tori, so it is an open question which instability
would prevail in PPI-unstable disks hosting small seed magnetic
fields.  In the context of neglecting the disk self-gravity, this
question was answered in~\cite{bugli_papaloizoupringle_2018}, where
simulations of accretion tori with weak toroidal seed fields (the
configuration with the slowest MRI growth rate) revealed that the MRI
overpowers the PPI, leading to a final state with no hint of the large
non-axisymmetries favored by the PPI. For disks that start out with
nearly axisymmetric configurations, this has effectively ruled out the
potential for the PPI to grow and produce detectable gravitational
waves, unless the initial field is very weak~\cite{nealon_papaloizoupringle_2018,bugli_papaloizoupringle_2018}.

However, when considering astrophysical processes that can produce
massive accretion disks around black holes it is plausible that disk
self-gravity could make a difference. Furthermore the dynamical events
that produce massive tori generically lead to configurations can be
significantly non-axisymmetric from the outset, e.g., tidal disruption
of a neutron star by a black hole. The disk self-gravity was not
treated in~\cite{bugli_papaloizoupringle_2018}, and no prior work has
determined what final configuration such disks will evolve to, or
whether such disks can retain significant non-axisymmetry long enough
to produce a detectable gravitational wave signal.

In this paper, we take the first step in exploring such
configurations.  We take advantage of the evolutions performed in our
previous study of the PPI~\cite{wessel_gravitational_2021}, in which
self-gravitating disks evolved into highly non-axisymmetric stable
configurations.  In this study, we resume two of these prior
simulations from near the time of maximum $m=1$ mode amplitude, but
with the introduction of small seed poloidal magnetic fields.  In this
way, these configurations can stand for a generic case of an initially
non-axisymmetric massive torus with weak initial magnetic fields. This
way we also give the PPI the best chance at surviving the magnetic
fields. In other words, if the PPI is suppressed by MRI when then PPI
is already saturated, then it is unlikely that PPI will be able to
dominate over the MRI in any other scenario.

In general we find that, for the configurations we simulated, the MRI
reduces the amplitude of the PPI-favored $m=1$ modes by a factor of 5
within a few orbits.  At late times, our simulations still show an
elevated $m=1$ amplitude in their settled states, which resuls in weak
but discernable GW signals.  Given that the resolution of our
simulations is limited we take our evolutions as characteristic of the
impact of magnetic fields on persistent non-axisymmetric modes in
accretion disks, and calculate the effect on GW detectability.  Since
our simulations end before the GW signals stop, we fit our extracted
GW waveforms to a simple model of the full duration signal, as we did
in~\cite{wessel_gravitational_2021}, and then exploit the scale
freedom in our evolution equations to scale the signals from our
simulations to a range of masses.  These are chosen to span a range
including BH-disks left over from BHNS mergers ($\sim 10 M_{\odot}$),
and supermassive star (SMS) collapse of ($\sim 1000 M_{\odot}$) and ($
\sim 2 \times 10^{5} M_{\odot}$).  We find that the maximum detection
luminosity distances are reduced by factors of 3.16 -- 6.67 for the
magnetized configurations.  This limits Cosmic Explorer (CE) to
detecting $10 M_{\odot}$ systems within 45Mpc, LISA to detecting $2
\times 10^{5} M_{\odot}$ systems within 2700Mpc, and DECIGO, which for
non-magnetized BH-disks could detect GWs from a $1000 M_{\odot}$
system out to $z \approx 5$, to within just $z \approx 2$.  Overall,
we find the optimistic model still sharply reduces the expectation of
detection for the magnetized disk configurations studied here, despite
the large head-start given to the PPI and the presence of strong
self-gravitation.

This paper is organized as follows.  In Section~\ref{section:methods}
we describe our methods. This covers the generation of initial data,
the evolution of hydrodynamic and magnetohydrodynamic disks, and the
definitions of all diagnostics we compute to track the behavior of our
BH disk systems.  In Section~\ref{section:results} we use these
diagnostics to quantify the dynamics of our disks and compare the new
magnetized evolutions to our prior un-magnetized ones.  We also
extract GW signals and repeat the analysis
of~\cite{wessel_gravitational_2021} to asses the effect magnetization
has on the maximum detection distance of such sources by LIGO, Cosmic
Explorer (CE), DECIGO, and LISA. We conclude in
Section~\ref{section:discuss} with a discussion of our results and
their broader implications for multimessenger astronomy.  In this work
we use geometrized units where \(G = c = 1\), except where stated
otherwise.  \(M\) designates the Christodoulou
mass~\cite{christodoulou_reversible_1970} of the central BH.

\section{Methods\label{section:methods}}
Here we summarize the methods used to produce initial data and evolve
it in this study.  These techniques have all been discussed in detail
elsewhere, and we provide appropriate references accordingly.  We also
describe in detail the diagnostics used in our analysis.  Analysis and
visualization of data produced by Cactus-based codes was carried out
with the help of the {\tt kuibit} python
module~\cite{bozzola_kuibit_2021}.

\subsection{Initial data\label{section:id}}
The initial configurations evolved in this study were previously
generated and evolved in~\cite{wessel_gravitational_2021}.  The only
significant difference between the two configurations is the spin
angular momentum of the central BH, which is zero in the case we refer
to as \So{}, but is parallel to the disk orbital angular momentum and
possesses a dimensionless magnitude of \(\chi = 0.7\) in the case we
refer to as \Su{}.  Using the techiques in
\cite{tsokaros_complete_2019}, the COCAL code solves the initial data
problem in general relativity (GR) for each configuration.  Starting
from an initially Kerr configuration, an elliptic form of the Einstein
equations are solved via the Komatsu-Eriguchi-Hachisu scheme
\cite{komatsu_rapidly_1989} as applied to BHs
\cite{tsokaros_numerical_2007}, producing spacetime and fluid initial
data for equilibrium non-magnetized BH-disk configurations with the
disk self-gravity included.  Both BH-disk configurations have equal BH
Christodoulou masses~\cite{christodoulou_reversible_1970}, \(M\), and
the inner edges of each disk are placed at the same radial coordinate,
which is well outside the respective innermost stable circular orbit
(ISCO).  Disk material is modeled as a perfect fluid with a polytropic
equation of state,
\begin{equation}\label{eq:polytrope}
  P = k \rho_{0}^{\Gamma},
\end{equation}
where \(\Gamma = 4/3\), \(\rho_{0}\) is the rest-mass density, and \(k\) is the polytropic constant, which scales out of the problem.

To be PPI-unstable, these disk configurations are designed to posses shallow spacific angular momentum profiles of the form,
\begin{equation}\label{eq:rot_law}
  j(\Omega) = A^{2} (B_{0} - \Omega),
\end{equation}
as in~\cite{tsokaros_complete_2019}, with $A=0.1$. $B_0$ and the other disk parameters are determined by an iteration proceedure that repeatedly generates configurations until finding ones that satisfy the requirements that the disk rest-masses be approximately $\sim0.1 M$.

\begin{figure}[!htb]
\includegraphics{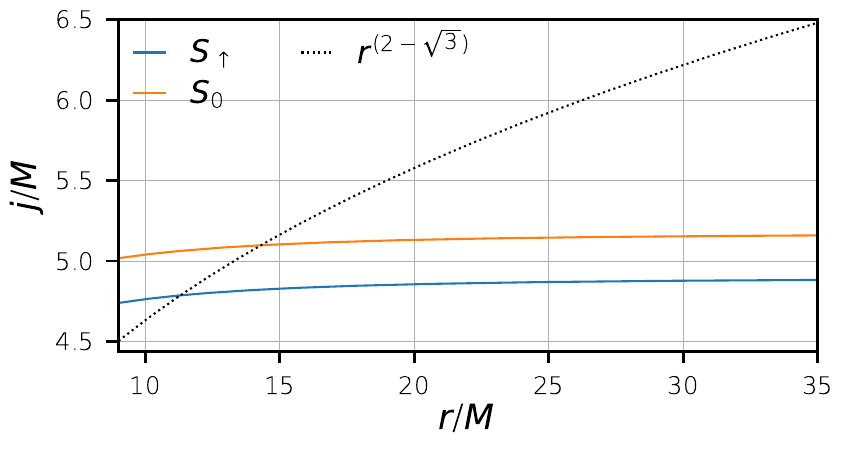}
\caption{The equatorial specific orbital angular momentum (\(j\)) as a function of radial coordinate (\(r\)). The black dotted line shows the steepest specific angular momentum profile that is PPI-unstable, configurations with steeper profiles will be stable to the PPI according to the criterion in Eq.~\ref{eq:PPI_criterion}.}\label{fig:specific_j}
\end{figure}

We also make sure that the disk rest-masses differ from one another by
less than $10\%$.  The result is self-gravitating disks on a
nearly-Kerr spacetime, where $j \sim r^{0.01}$, as shown in Figure
\ref{fig:specific_j}, which corresponds to $q \approx 2$ thereby
satisfying the criteria of Eq. \ref{eq:PPI_criterion} for
PPI-susceptability\footnote{The disks are not gravitationally unstable: the specific angular momentum profiles are steep enough that the Toomre stability criterion\cite{toomre_gravitational_1964-1} is robustly satisfied, even ignoring the stablizing effect of finite thickness.}.
The parameters for both configurations are presented in Table~\ref{tab:initial_data}.

\renewcommand{\tabcolsep}{1pt}
\begin{table}[!hbt]
\caption{Parameters of initial data.
All masses, distances, and timescales expressed in terms of BH Christodolou mass \cite{christodoulou_reversible_1970} \(M\).
From left to right we have: the dimensionless spin parameter of central BH (\(\chi\));
the ADM mass of the spacetime (\(M_{\mathrm{ADM}}\));
coordinate radii of the inner-most stable circular orbit (\(r_\mathrm{ISCO}\))
inner edge of the disk (\(r_\mathrm{inner}\)),
maximum density point of the disk (\(r_c\)),
outer edge of the disk (\(r_\mathrm{outer}\));
the disk orbital period at \(r_{c}\) (\(t_{\mathrm{orb}}\));
and the rest-mass of the disk (\(M_\mathrm{disk}\)).}\label{tab:initial_data} \def\arraystretch{1.5}
\begin{tabular}{ccccccccc}
  \hline\hline
Model \ & \bfseries \(\chi\) & \bfseries \(\frac{M_\mathrm{ADM}}{M}\) & \bfseries \(\frac{r_\mathrm{ISCO}}{M}\) & \bfseries \(\frac{r_\mathrm{Inner}}{M}\) & \bfseries \(\frac{r_c}{M}\) & \bfseries \(\frac{r_\mathrm{Outer}}{M}\) & \bfseries {\(\frac{t_{\mathrm{orb}}}{M}\)} & \bfseries {\(\frac{M_\mathrm{disk}}{M}\)} \\ 
\hline \Su{}\ & 0.7 & 1.13 & 3.39 & 9.00 & 15.6 & 31.7 & 390 & 0.12 \\
 \So{}\ & 0.0 & 1.14 & 6.00 & 9.00 & 16.9 & 35.0 & 427 & 0.135 \\
 \hline\hline
\end{tabular}
\end{table}

The COCAL code solves for initial data configurations on a polar coordinate system in which the interior of the BH is excised.
In order to evolve this data with a moving-puncture scheme as described in Section~\ref{section:non-mag_ev}, it is necessary to provide initial data for the excised region as well, since it includes parts of the BH interior that are still covered by the puncture coordinates.
Fortunately, the additional initial data can be constraint-violating without influencing the BH exterior as long as it is sufficiently smooth, and so we use the so-called ``smooth junk-filling'' technique to fill the BH interior~\cite{etienne_filling_2007}.

\subsection{Evolution grid structure\label{section:grid_structure}}
For time evolution the initial data is interpolated from the spherical
coordinates used by COCAL onto cartesian grids.  Our grids are a
nested hiearchy of concentric cubes of half-side lengths which are
powers of 2 starting at the finest level \(2.19 M\) and extending to
\(2250 M\) for a total of 11 levels of refinement, with each sucessive
level having half the resolution of the coarsest one that it contains.
This allows the stringent requirements for resolving the event horizon
near the BH to be met, as well as the more relaxed requirements for
resolving disk dynamics, while avoiding excessive resource use in the
outlying regions of the domain where there is little spatial or
temporal variation.

In our experience, it is not uncommon for evolutions of the Baumgarte-Shapiro-Shibata-Nakamura (BSSN) formulation of general relativity in our code to slowly accumulate secular drifts of the coordinates, which, over long evolutions, result in the center of mass of the system gradually meandering. Indeed, we encounter just such a drift in these simulations as well as those of~\cite{wessel_gravitational_2021}.
The exact cause of this phenomena in our code has not been closely studied.
In our evolution of the shift we use up-winding stencils which could potentially be sensitive to a minor numerical instability leading to an error build-up which could cause this coordinate motion.
While the drift starts too early for any physical causal influence from the boundary to have propagated into center of the domain, BSSN exhibits some gauge modes which are super-luminal, so this could also be a boundary effect.
There are other concievable explanations, however, regardless of the cause, the secular coordinate drift only impacts the coordinate gauge and does not directly influence the physical degrees of freedom of the system.

On the other hand, the coordinate drift is a hazard to the mesh-refinement scheme, since it might cause the BH to stray accross refinement boundaries and become under-resolved.
To prevent this, the simulation is periodically re-gridded every $2.5M$,
with all data interpolated onto a new set of grids centered on the
current BH apparent horizon centroid\footnote{Although the grid structure is moved, we do not alter the coordinates themselves, so this proceedure does not repeatedly move the BH to the center of the coordinate system.}.
Table~\ref{tab:grid_info} lists the grid parameters. The jump in size between the \(n=3\) and \(n=4\) grids is to ensure that the disk is well-resolved and to avoid a mesh-refinement boundary intersecting with the bulk of the disk, which could introduce artifacts.
This compares favorably to the grids used to study the PPI in~\cite{korobkin_stability_2011, kiuchi_gravitational_2011}, and in our prior study~\cite{wessel_gravitational_2021} this exact structure was used and shown to sucessfully capture the dynamics of the PPI.

\renewcommand{\tabcolsep}{10pt}
\begin{table}[!hbt]
\caption{Parameters of the nested cartesian grids. From left to right: the refinement level (\(n\)), the box half-side length (or ``minimum radius'') for the grid of level \(n\) (\(r_{\mathrm{box}}(n)\)), the grid spacing for the grid of level \(n\) (\(\Delta(n)\)). The grids are all perfect cubes initially centered at the origin but periodically updated to follow the BH centroid. Grid spacings are equal to \(\Delta(n)\) in the \(x\), \(y\), and \(z\) direcitons.}\label{tab:grid_info} \def\arraystretch{1.5}
\begin{tabular}{ccc}
  \hline\hline
  \(n\) & \(r_\mathrm{box}(n)\) & \(\Delta(n)\) \\
  \hline
  1 & \(2.19 M\) & \(M/25.6\) \\
  2-3 & \(2^{(n-1)} r_\mathrm{box}(1)\) & \(2^{(n-1)} \Delta(1)\) \\
  4-11 & \(2^n r_\mathrm{box}(1)\) & \(2^{(n-1)} \Delta(1)\) \\
  \hline\hline
\end{tabular}
\end{table}

\subsection{Non-magnetized evolution\label{section:non-mag_ev}}
Initially, we evolve our disks purely hydrodynamically.  These are the
same evolutions reported previously
in~\cite{wessel_gravitational_2021} for the \Su~and \So~cases.  We use
the {\tt Illinois GRMHD} adaptive-mesh-refinement (AMR)
code~\cite{duez_relativistic_2005,etienne_relativistic_2010,etienne_general_2012},
which was built within the open-source Cactus-Carpet
infrastructure~\cite{noauthor_cactus_nodate-1,schnetter_evolutions_2004}
and is itself the basis of the publicly available counterpart in the
Einstein Toolkit~\cite{etienne_illinoisgrmhd_2015}.  This code evolves
the metric via the BSSN formulation of the Einstein
equations~\cite{shibata_evolution_1995,baumgarte_numerical_1998}.
Moving puncture gauge conditions are
used~\cite{campanelli_accurate_2006,baker_gravitational-wave_2006}
with \(1+\log\) time slicing~\cite{bona_new_1995} and the ``gamma
driver'' shift condition~\cite{alcubierre_gauge_2003} cast into a
first-order form as in Eq. 18 of~\cite{hinder_error-analysis_2013}.

The disk material is treated as a perfect fluid with a \(\Gamma\)-law equation of state (EOS),
\begin{equation}\label{eq:eos}
  P = (\Gamma - 1)\rho_{0}\epsilon,
\end{equation}
where \(\epsilon\) is the internal specific energy, and \(\Gamma =
4/3\) as before.  This fluid model is consistent with the polytropic
fluid model used in the initial data (Eq. \ref{eq:polytrope}) when the
disk material is isentropic, as it is assumed to be in the initial
data.  The choice of \(\Gamma = 4/3\) is realistic for an
optically-thick radiation-pressure dominated gas, and allows also close
comparison to the work of~\cite{bugli_papaloizoupringle_2018}.

\subsection{Magnetic field insertion\label{section:magfield}}
Using checkpoints of our original hydrodynamic evolutions we resume
our simulations from shortly after the PPI growth saturates -- when the
$m=1$ PPI non-axisymmetic mode is near its maximum, and after a
short period of hydrodynamic evolution when the
time of all refinement levels is the same, we seed an initial magnetic field
to render the disk MRI-unstable.

The vector potential and magnetic field, here defined with respect to coordinate-stationary observers, are $A_{i}$ and $B_{i}$, respectively, and they are related by,
\begin{equation}\label{eq:vector_potential_def}
  B_{i} = \epsilon^{i j k}\partial_{j}A_{k},
\end{equation}
where $\epsilon^{i j k}$ is the Levi-Civita tensor associated with spatial slice metric $\gamma_{i j}$.
In terms of the vector potential, our initial field is,
\begin{equation}\label{eq:seed_field}
  A_{\phi} = A_{b}\max{(P-P_{\mathrm{cut}}, 0)}.
\end{equation}
$A_{b}$ is a strength parameter of the field, which indirectly determines the ratio of magnetic to gass pressure at the disk center.
The cutoff term $P_\mathrm{cut}$ is chosen to be 10\% of the maximum pressure to ensure that the field is supported only within the high-density region of the disk, and the purely $\phi$-directed vector potential results in a $B$-field that is initially poloidal. Critically, the configurations we study here have pressure below $P_{\mathrm{cut}}$ all along the azimuthal axis, and in our hydrodynamics scheme all shocks are smoothed over a few grid cells, so this choice of vector potential has no discontinuities. Additionally, although the $\max$ function is not formally differentiable at one point, this is not a problem for derivatives calculated by finite difference methods.

To choose the field strength $A_{b}$, we consider the fact that in order for the disk to be MRI-unstable, $\lambda_{\mathrm{MRI}}$ must fit within the vertical height of the high-density region of the disk, otherwise the growth of the MRI modes would be incompatable with the effective boundary conditions at the disk edge.
The MRI wavelength is,
\begin{equation}\label{eq:mri_wavelength}
  \lambda_{\mathrm{MRI}} \equiv 2\pi\frac{|v_A|}{|\Omega|},
\end{equation}
where $\Omega$ is the orbital frequency for local material.
The Alfvén velocity is,
\begin{equation}\label{eq:alfven_vel}
  v_{A} = \sqrt{b^2/(b^2 + \rho_0(1 + \epsilon) + P)},
\end{equation}
where $b^{i} = B^{i}_{u} / \sqrt{4 \pi}$ and $B^{i}_{(u)}$ is the magnetic field observed in the fluid rest-frame~\cite{etienne_relativistic_2010}.
In the limit of a weak field \(b^{2} \ll \rho_0(1 + \epsilon) + P\), so
\(v_{A} \propto |b| \) in general, which implies \(\lambda_{\mathrm{MRI}} \propto |b|\).
Therefore, the constraint that the MRI mode fit within the disk provides an upper
bound on the value of $b^{2}$, or equivalently $B^2$, and thus on the field strength parameter
$A_{b}$.

For the lower limit on $A_{b}$, we follow the empirical guidance of~\cite{hawley_assessing_2011}, who found that an average
of \(\lambda_{\mathrm{MRI}}/\Delta(n) \approx 10\) in the vertical
direction and \(\approx 20\) in the toroidal direction was sufficient
to resolve the MRI turbulent state. In our case, the dense region of the disks sits between about $10 M$ and $35 M$, and so $\Delta(4)$ from Table~\ref{tab:grid_info} is the grid-spacing that determines the minimum resolvable $\lambda_{\mathrm{MRI}}$.

\begin{figure}[!htb]
\includegraphics{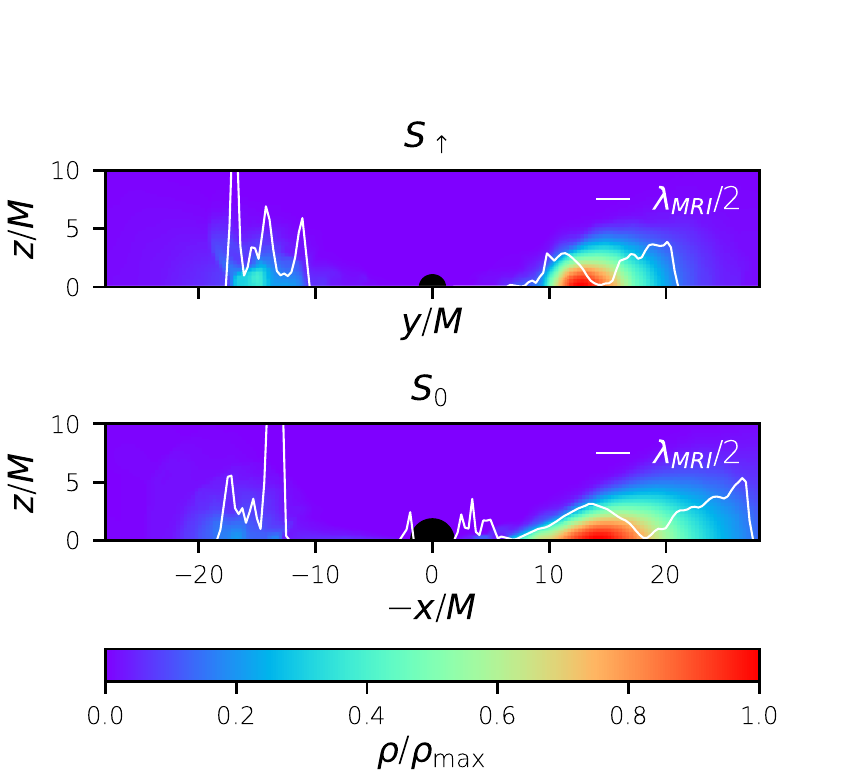}
\caption{Cross-sections of the baryon rest-mass density in both accretion disks at the time of field insertion, superimposed with a plot of half the vertical MRI wavelength ($\lambda_{\mathrm{MRI}}$) at the midplane. In both cases, the MRI modes fit within the disks on the overdense side, while the MRI wavelength on the rarified side is too large for modes to grow there. For \SuM~the $yz$ cross section is used, while for \SoM~the $xz$ cross section is used, with the $x$-axis flipped to align the high-density sides of the disks.}\label{fig:lambda_MRI_overlay}
\end{figure}

For both the disks we evolved, we seeded fields with parameters that met all these criteria.
Figure~\ref{fig:lambda_MRI_overlay} compares the disk rest mass configuration to the vertical MRI wavelength at the time of field insertion.

\begin{figure}[!htb]
\includegraphics{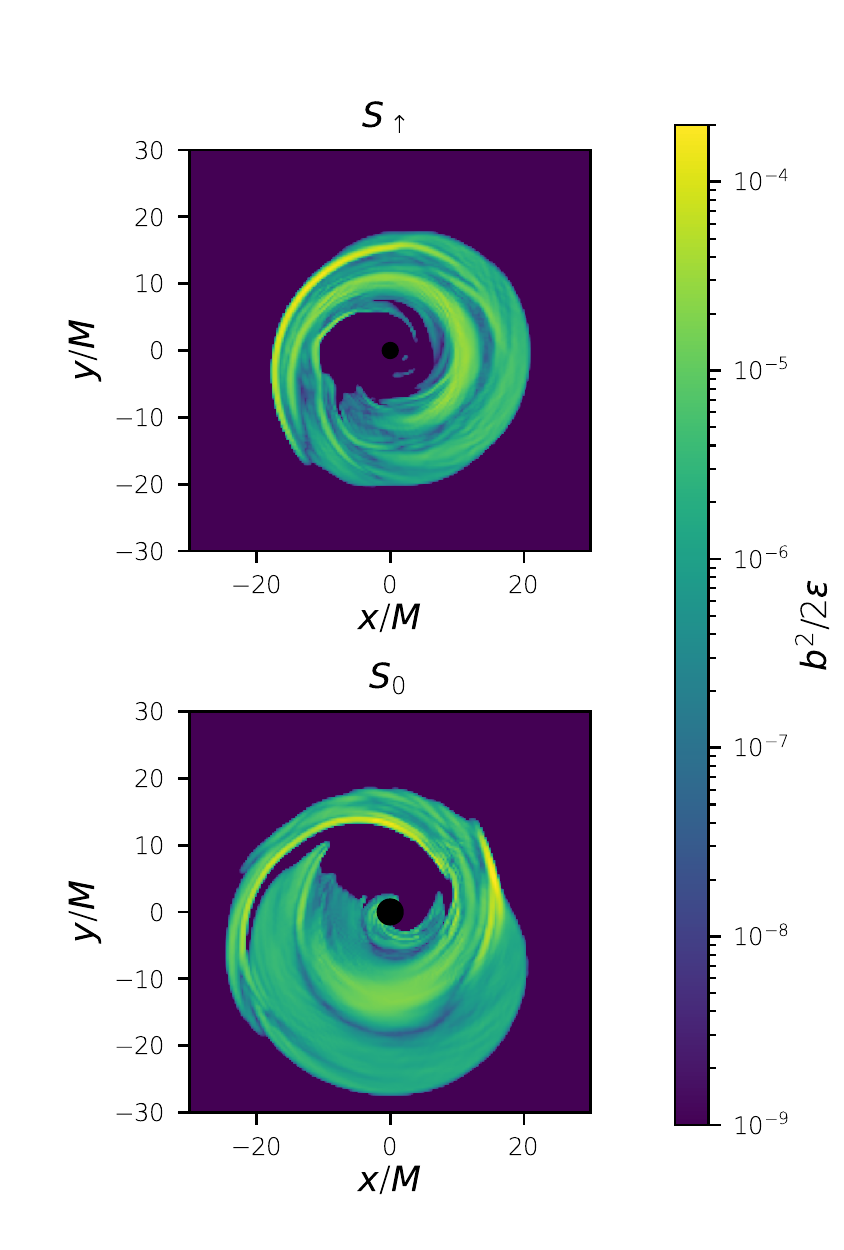}
\caption{Ratio of the magnetic energy density ($b^{2}/2$) to the gas internal energy ($\epsilon$) in the midplane of the disks at the moment of field insertion, on a log scale. Nowhere is the ratio greater than $10^{-4}$, so the fields are extremely weak and do not disrupt the gas or introduce non-negligable constraint violations.}\label{fig:mag_energy_ratio}
\end{figure}

We can also compare the magnetic energy of the seed field to the internal energy of the gas at the moment of field insertion, which is shown in Figure~\ref{fig:mag_energy_ratio}. As can be seen from the figure, the inserted field is weak in the sense that the energy is still dominated by the gas. This means that the gas will not be significantly disrupted by the field, and that the resulting magnetized configuration is still constraint-satisfying to within numerical fidelity. Thus, we can think of the moment of field insertion as specifying a new initial configuration, which can be evolved and meaningfully compared to our previous non-magnetized evolutions.

\subsection{Magnetized evolution}
With magnetic fields now inserted, we continue the evolution using the
{\tt Illinois GRMHD} AMR code (as described in
Section~\ref{section:non-mag_ev}) adopting ideal magnetohydrodynamics.
The magnetic induction equation is evolved using a vector potential
method to ensure perfect satisfaction of the \(\nabla \cdot B = 0\)
condition~\cite{etienne_relativistic_2010}.  This code has previously
been used for numerous studies of magnetohydrodynamics in extreme
dynamical spacetimes, such as magnetized BH-NS
mergers~\cite{etienne_general_2012,paschalidis_relativistic_2015}, MRI-driven jet-launching from BNS
merger remnant disks~\cite{paschalidis_general_2017}, MRI-driven
accretion disks around binary black
holes~\cite{gold_accretion_2014,paschalidis_minidisk_2021}, etc. For
our electromagnetic gauge choice, we adopt the generalized Lorenz
gauge condition developed in~\cite{farris_binary_2012}, which avoids
the development of spurious magnetic fields across the AMR levels,
setting the generalized Lorenz gauge damping parameter to $\xi =
0.5/M$ for the non-spinning case and $\xi = 0.65/M$ for the spinning
case. A slightly higher value was used in the spinning case because an early test run suffered an instability triggered by magnetic flux build-up on the refinement boundary nearest the BH.
Since the dynamics of magnetic fields near the spinning BH are different than for the non-spinning case, and the curvature at the horizon is slightly greater, the need for a slightly different level of damping was not totally unexpected. As this parameter only impacts the gauge evolution, it should have no effect on the physical dynamics of the simulations besides keeping them numerically stable.
Separately from the second-order dissapation in the generalized Lorenz gauge, we apply higher-order Kreiss-Oliger dissipation to the vector potential and fluid quantities inside the apparent horizon. This suppreses any poorly-resolved short-wavelength modes that might appear there, which could otherwise potentially leak out of the horizon due to finite-difference truncation error.

\subsection{Diagnostics\label{section:diagnostics}}
To ensure the quality of the spacetime evolution, we monitor the normalized Hamiltonian and momentum constraints as defined in \cite{etienne_fully_2008} (Eqs.~(40)-(43)).
The output of these diagnostics is discussed in the Appendix~\ref{appendix:constraints}.

The evolution of non-axisymmetric density modes is tracked in two ways.
Using 3D grids of the rest-mass density \(\rho_{0}\) output every $80M$ during the simulation, we define a simple non-axisymmetric \textit{mode amplitude} diagnostic,
\begin{equation}\label{eq:mode_def_new}
  C_m = \int \mathrm{d}^3x \rho_0 e^{im\phi}.
\end{equation}
Where $\phi$ is the azimuthal coordinate of a polar coordinate system centered on the BH centroid.
Because the BH dominates the mass of the system, the center of mass is always near the origin of this coordinate system, so this diagnostic compensates for the slow secular drifting of our coordinates described in Section~\ref{section:grid_structure}, which would otherwise induce mode-mixing between different $m$ and undermine the interpretation of our results.

The definition of mode amplitudes used here is slightly different from that used in prior studies~\cite{paschalidis_one-arm_2015,east_equation_2016,east_relativistic_2016,wessel_gravitational_2021},
\begin{equation}\label{eq:mode_def_old}
C^{\mathrm{old}}_m = \int{\sqrt{-g}\mathrm{d}^3x u^0\rho_0 e^{im\phi}},
\end{equation}
where \(g\) is the determinant of the full space-time metric and \(u^{0}\) is the time component of the fluid 4-velocity.
Unfortunately, while we did extract Eq.~\ref{eq:mode_def_old} every $10M$, the routine that computed it was not set up to evaluate the integral in a BH-tracking coordinate system, so at late times when the center of mass is far from the coordinate origin the quality of this diagnostic suffers noticably.
In practice, Eq.~\ref{eq:mode_def_new} agrees closely with Eq.~\ref{eq:mode_def_old} during the times when the system center of mass has not yet drifted too far from the origin (despite the difference in integration measure), and although the frequency of sampling is lower, Eq.~\ref{eq:mode_def_new} is still computed multiple times per orbit, capturing dynamics down to the orbital freqency.

While our mode amplitudes can be informative, they do not exactly match the diagnostics used by~\cite{bugli_papaloizoupringle_2018}, who instead extract the \textit{mode power}, normalized with respect to the \(m=0\) mode,
\begin{equation}\label{eq:mode_power}
  \frac{P_{m}}{P_{0}} = \frac{\int^{r_{\mathrm{out}}}_{r_\mathrm{in}}{\int^{\pi}_{0}{ \left|\int^{2\pi}_{0}{\rho_{0} e^{im \phi} \mathrm{d}\phi } \right|^{2} \sqrt{\gamma} r^{2}\cos{\theta} \mathrm{d}\theta}\mathrm{d}r}}{\int^{r_{\mathrm{out}}}_{r_\mathrm{in}}{\int^{\pi}_{0}{ \left| \int^{2\pi}_{0}{\rho_{0} \mathrm{d}\phi} \right|^{2} \sqrt{\gamma} r^{2}\cos{\theta} \mathrm{d}\theta}\mathrm{d}r}},
\end{equation}
where $r, \theta, \phi$ are polar coordinates with their origin at the BH centroid (in \cite{bugli_papaloizoupringle_2018} this coincides with the COM), and \(\gamma\) is the determinant of the spatial metric in cartesian coordinates.
To compute this measure of the \(m\)-mode strengths for our own evolutions we do a linear interpolation of our cartesian 3D grid data onto a set of grid points in a polar coordinate system centered on the BH.
The polar grid has points at 150 evenly spaced radial coordinate values spanning \(M \leq r \leq 100 M\), 100 evenly spaced $\theta$ values between $- \pi/2$ and $\pi/2$, and 200 evenly spaced $\phi$ values between 0 and $2\pi$. Once the quantities have be interpolated onto this grid we evaluate the integrals via the trapezoidal method.

\begin{figure}[!htb]
\includegraphics{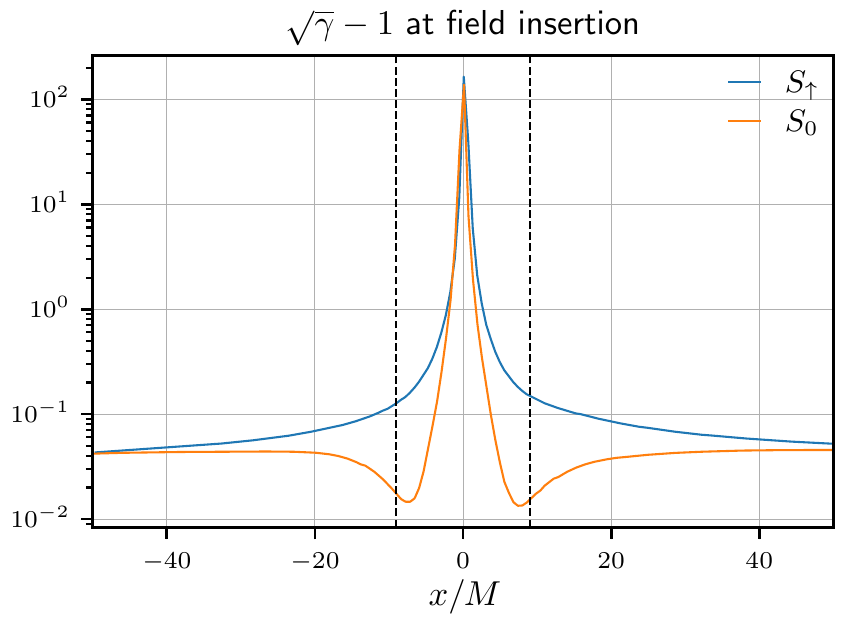}
\caption{The difference between the square root of the spatial metric determinant ($\sqrt{\gamma}$) and the cartesian flat-space value (1), on a log scale, for both configurations at the time of field insertion. The dashed black lines show the $r=9M$ coordinate radius. Because most disk material lies outside of $9M$ in our evolutions, approximating $\sqrt{\gamma}\approx 1$ is appropriate when evaluating diagnostic integrals for these simulations.}\label{fig:sqrt_gamma_compare}
\end{figure}

The spatial metric determinant \(\gamma\) was not regularly saved in our evolutions; however it was recorded at a few key times, allowing us to determine that by the time of field insertion \(\gamma\) has settled to a static state. Figure~\ref{fig:sqrt_gamma_compare} shows the difference between $\sqrt{\gamma}$ and 1 along a line through the BH at the time of field insertion, for both configurations. The initial inner edge of the disks is at $r = 9M$ in both cases, and we found that the integrals are dominated by material further than $9M$ from the BHs throuought the evolutions as well. Outside this distance, it can be seen that $\sqrt{\gamma}$ differs from 1 (the flat-space cartesian value) at the $\sim10\%$ level.

Therefore, even though we lack the true values of $\sqrt{\gamma}$ at all times, we are able to approximate $\sqrt{\gamma} \approx 1$ when computing Eq.~\ref{eq:mode_power} and still make rough comparisons with
the results of~\cite{bugli_papaloizoupringle_2018}, since our values are not likely to me more than a few percent different on account of the approximation.

To keep track of the magnetic state of the disk, we output 3D grids of $B_{i}$ every $80M$ as well.
From these data, we can calculate many useful diagnostics.
As in~\cite{hawley_assessing_2011,hawley_testing_2013}, we also compute the \(\rho_{0}\)-weighted average \(B\)-field square magnitude in the $i=\hat{\phi}, R, z$ directions as,
\begin{equation}\label{eq:B_avg}
  \langle B_{i}^{2} \rangle \equiv \frac{\int{B^{2}_{i} \rho_{0} \mathrm{d}^{3}x}}{\int{\rho_{0} \mathrm{d}^{3}x}},
\end{equation}
where $R$ is the radial component of the cylindrical coordinate basis (rather than the $r$ component, which points in the sperical radial direction). The index $\hat{\phi}$ is meant to indicate the \textit{orthonormal} $\phi$ basis component.
Therefore $B_{\hat{\phi}}$ will have units of $1/M$ just like $B_{z}$ and $B_{R}$.
This convention is used by~\cite{hawley_assessing_2011,hawley_testing_2013}.

Gravitational waveforms were extracted using the $\Psi_{4}$ Weyl
scalar of the Newman-Penrose formalism on the surfaces of concentric
coordinate spherical shells centered on the origin.  The extraction
radii were chosen to be distant enough from the source to be safely in
the far-field, while being near enough that the grids still resolved
the GW wavelengths well. This was achieved by choosing extraction
radii at which the signals fall-off as $1/r$, a behavior from which the
signals deviate if $r$ is too close to or too far from the BH (and hence in an
under-resolved region). The complex value of $\Psi_{4}$ over each
extraction sphere was decomposed into $s=-2$ spin-weighted spherical
harmonics for all modes with $\ell \le 3$.  The strain $h$ is the
double time integral of $\Psi_{4}$, and was computed via the fixed
frequency integration technique of~\cite{reisswig_notes_2011}, with
the real and complex parts providing the $+$ and $\times$ GW
polarizations in the natural basis: $h = h_{+} + i h_{\times}$.

\subsection{Resolution dependence\label{section:resolution}}
We lacked the computational resources to conduct a formal resolution study as a part of this work.
In addition to the $O(\Delta^{-4})$ scaling of resource requirements typical of 3D time evolutions, the resources needed to perform a resolution study in our case were further inflated by the requirement that we re-evolve the hydrodynamic PPI to the point of saturation before inserting magnetic fields. Therefore, assessing the impact of resolution on the turbulent magnetic field state would have been especially expensive.

Nonetheless, we can indirectly asses the impact resolution is likely to have given what is known of our code's properties as well as the results of prior studies evolving similar systems.

The {\tt Illinois GRMHD} AMR code is designed to be second-order convergent for hydrodynamic quantities, and tests have shown it to be sucessfull~\cite{duez_relativistic_2005,etienne_relativistic_2010,etienne_general_2012,etienne_illinoisgrmhd_2015}. In prior work it has performed as expected in similar situations, both for pure relativistic hydrodynamics and magnetohydrodynamcs~\cite{espino_dynamical_2019,paschalidis_minidisk_2021,gold_accretion_2014}.

The resolution of our grids, as detailed in Section
\ref{section:grid_structure}, easily matches or exceeds that of prior
numerical evolutions of the hydrodynamic PPI in general
relativity~\cite{kiuchi_gravitational_2011,korobkin_stability_2011}.
In addition, as reported in~\cite{wessel_gravitational_2021}, we
performed a set of incomplete runs with a different grid strucutre,
where a refinement boundary crossed into the dense regions of the
disk, and most of the disk evolved with only half the resolution.  In
these runs the early hydrodynamic evolution of the PPI was nearly
identical to our production runs with the higher resolution grid, so
we are confident that the spacetime and hydrodynamic evolutions are
adequately resolved.

For magnetized accretion tori, capturing the turbulent state induced
by the MRI is crucial for sustaining the disk
dynamo~\cite{balbus_instability_1998,blaes_general_2014}.  This
turbulence is challenging to resolve, requiring many decades of length
scales to be captured in order to reach full convergence.  What level
of resolution is adequate to obtain realistic bulk behavior for
MRI-driven accretion disks in GRMHD has been the subject of numerous
studies~\cite{hawley_assessing_2011,hawley_testing_2013,ryan_resolution_2017}.
While there is reason to question whether any global accretion disk
simulations are yet fully convergent~\cite{ryan_resolution_2017},
simulations where $\lambda_{\mathrm{MRI}}/\Delta \approx 10$ in the
vertical direction and $\approx 20$ in the azimuthal direction in the
high-density regions of the disk are thought to capture MRI-driven
turbulence decently~\cite{hawley_testing_2013}.  As mentioned
previously in Section~\ref{section:magfield}, our disks meet this
criteria upon field insertion, and for a period of time
after. Therefore, while the qualitative behavior of our results is on
solid footing, making quantitative error estimates requires a
resolution study which goes beyond the scope of the current work.

\section{Results\label{section:results}}
\begin{figure*}[!tb]
  \includegraphics{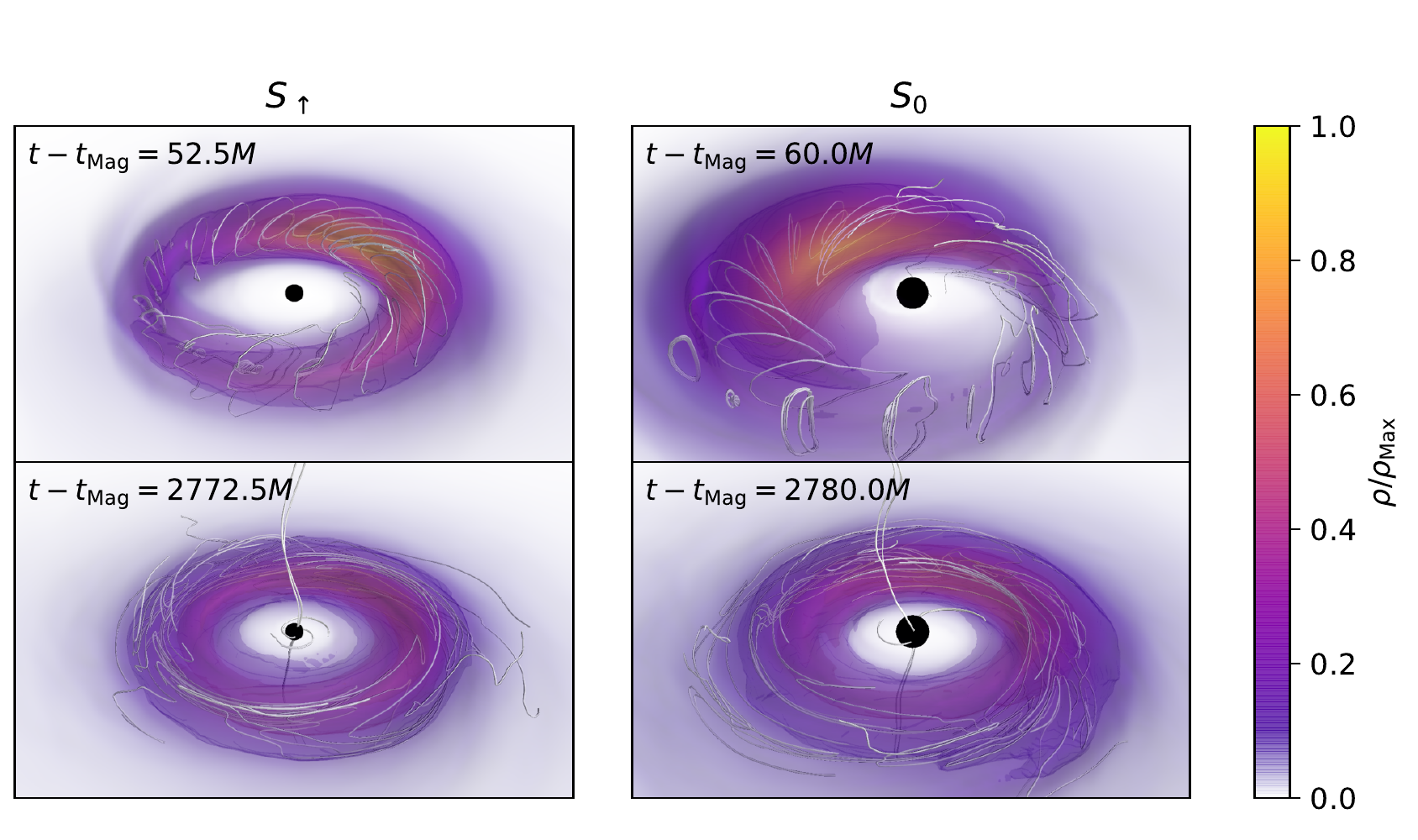}
  \caption{Perspective renders of both disks at different times after magnetic field insertion. The left column shows \SuM{}, the right shows \SoM{}; the top panels show the disks shortly after magnetic field insertion, the bottom panels show the disks later on. Volumetric shading indicates baryon number density, and the surfaces within the disks are isodensity contours to further emphasize the disk structure. Magnetic field lines are traced by the shaded white filaments, the black hole apparent horizon is black, and in all panels the disks are viewed from a coordinate distance of $70M$ from the BH centroid at an angle of $35^{\circ}$ above the $xy$-plane. The difference in size of the apparent horizons is a known effect of the chosen coordinates around spinning black holes. While the $m=1$ non-axisymmetries are present with a large amplitude shortly after magnetic field insertion, at later times the action of the MRI has significantly reduced their amplitudes, while the field configuration has become much more turbulent and toroidal, as expected from the MRI, and even features some magnetic field lines radiating from the BH pols in a manner reminiscent of jets (although we do not study jets here, and these features are seen in both the spinning and non-spinning cases).}\label{fig:3d_comparison}
\end{figure*}

The two BH-disk systems we studied, the spinning BH case \Su~(\(\chi =
0.7\)) and the non-spinning case \So, were previously simulated in
full GR without magnetic fields~\cite{wessel_gravitational_2021}.  Of
the three configurations in our previous study, these were the two
that produced the most powerful and sustained GW signals.  In this
study we do not evolve the configurations from their initial
axisymmetric states, but instead we resume the evolutions near the PPI
saturation time, when the \(m=1\) mode is at its maximum, and insert a
seed poloidal magnetic field to study its effect on the subsequent
dynamics. As can be seen from the 3D renders in Figure~\ref{fig:3d_comparison}, the $m=1$ non-axisymmetries are reduced significantly by the introduction of magnetic fields and the growth of the MRI. In the following sections, we make this assesment more quantitative, and determine what it implies for potential detectability.

\subsection{MRI growth and saturation\label{section:MRI_growth}}
\begin{figure}[!htb]
\includegraphics{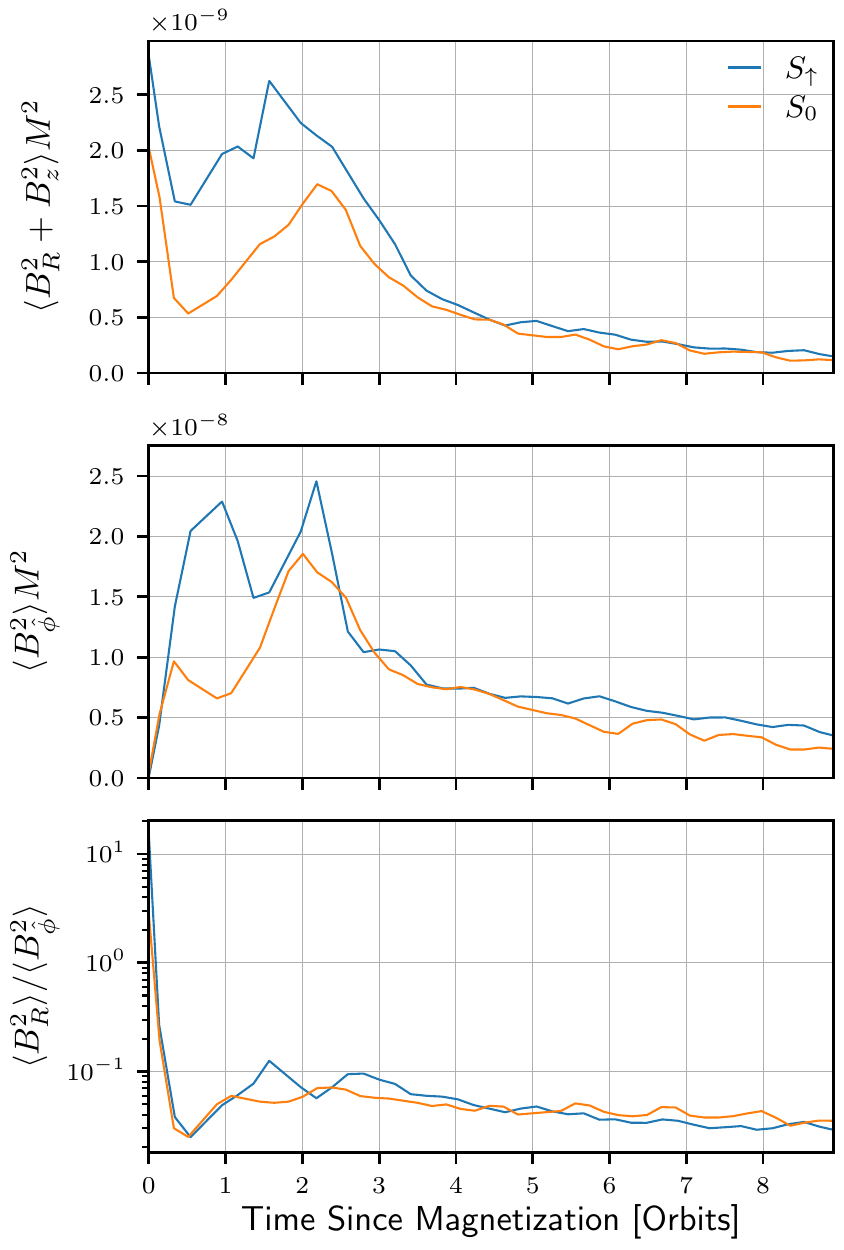}
\caption{The density-weighted square-averages of magnetic field components (as defined by Eq.~\ref{eq:B_avg}). Top: Sum of the non-azimuthal field components $R$ and $z$. Middle: Azimuthal field component $\hat{\phi}$. Bottom: Ratio of the square-averages of the $R$ and $\hat{\phi}$ components, on a log-scale.}\label{fig:B_strength}
\end{figure}

We can track the MRI growth and saturation through averages of the magnetic field components.
Figure~\ref{fig:B_strength}, show time-series of the average squared values of $B$-field components, weighted by rest-mass density $\rho_{0}$ (Eq.~\ref{eq:B_avg}).
The dynamics in both the $S_{\uparrow}$ and $S_{0}$ are similar. Initially, the $R$ and $z$ components of the field have a finite value, while the toroidal component is near zero. Subsequently differential rotation rotates non-toroidal components into the toroidal direction, reducing the former and increasing the latter.
Quickly the decline of the non-toroidal components halts and reverses, consistent with the expected behavior of the MRI where back-reaction of the increasing field line tension on the orbiting fluid leads to run-away growth of the field strength (see Section IV.B of~\cite{balbus_instability_1998} for an intuitive treatment).
Within two orbits (as measured by the orbital period of the maximum density material in the inital data) the MRI appears to have saturated, with both the $S_{\uparrow}$ and $S_{0}$ simulations reaching their maximum magnetic strengths.

After this initial growth plateaus, the magnitudes of the magnetic
field averages begin to decay.  This is commonly seen in accretion
disk simulations, see for example Figure~1b
of~\cite{hawley_testing_2013} (although note our initial field
configuration is different), and is associated with a general decline
in magnetic energy~\cite{hawley_assessing_2011,hawley_testing_2013}.
This decline could be due to the depletion of the disk.  We can look
at diagnostics insensitive to the overall flux: the bottom panel of
Figure~\ref{fig:B_strength} shows the
ratio of the $R$ and $\hat{\phi}$ $B$-field components.  The value this
ratio relaxes to, about 0.05, is typical of slightly under-resolved
MRI turbulence seen in previous studies (Figure~4
of~\cite{hawley_assessing_2011} and Figure~5
of~\cite{hawley_testing_2013}).

\subsection{Accretion rate\label{section:accretion}}
\begin{figure}[!htb]
\includegraphics{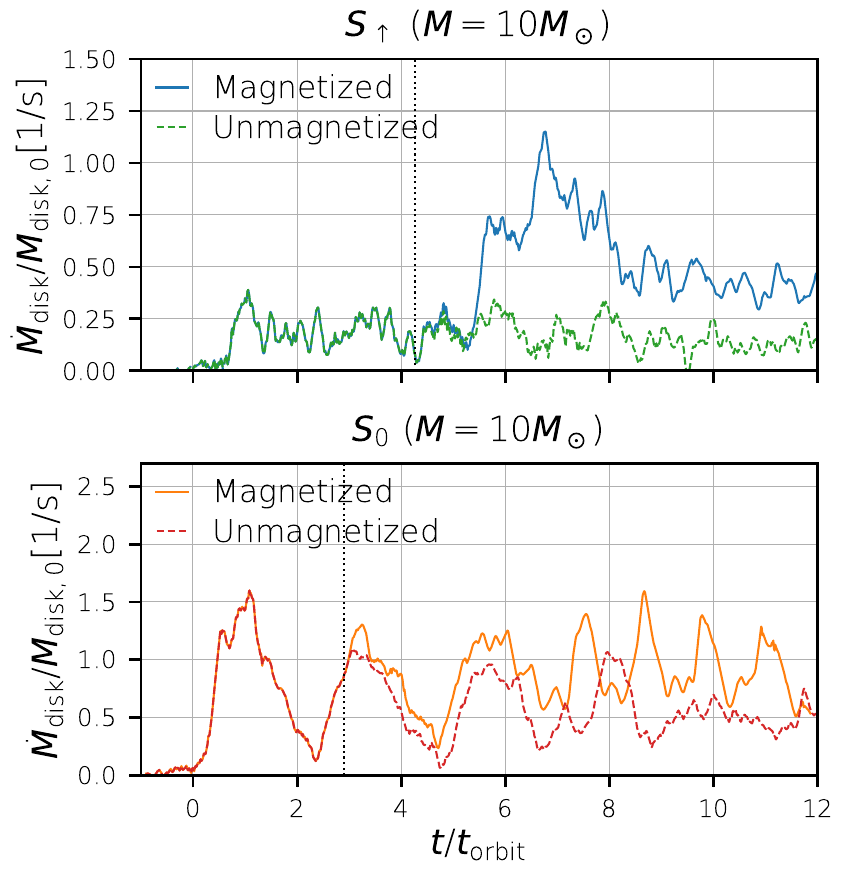}
\caption{Comparison of accretion rates between the non-magnetized BH-disk systems (dashed lines) and same simulations re-run with magnetic fields seeded at the indicated times (solid lines). The dotted vertical lines indicate the time of magnetic field insertion.}\label{fig:accretion}
\end{figure}
The ultimate source of power for BH-disk systems is the release of gravitational binding energy as material is accreted onto the central black hole.
The accretion rate therefore determines the maximum power that such systems can emit in any form, including EM and GW emission.
In combination with the total mass of the tori, the rate of accretion also determines the lifetime of these objects, and therefore bounds the lifetimes of any potential observable signatures.
The conservation of net angular momentum means that accretion can only be enabled by transport of angular momentum from inner to outer disk regions, and the PPI and MRI instabilities have long been known to produce conditions that can make this transport efficient in astrophysical disks \cite{balbus_instability_1998}.
In Figure~\ref{fig:accretion}, the accretion rate is tracked as a function of time for the original hydrodynamic evolutions--subject only to the PPI--and the new MHD evolutions, where both the PPI and MRI are present.
We see that the magnetized disks accrete faster than their non-magnetized counterparts.
The main impact of the increased accretion rate for this study will be the resulting shortening of the disk lifetime, which we will account for when placing an upper bound on the GW detectability.

\subsection{Decay of low-$m$ density modes\label{section:mode_decay}}
We can directly track the mode amplitudes via the diagnostic described in~\ref{section:diagnostics} (Eq.~\ref{eq:mode_def_new}).
\begin{figure}[!htb]
\includegraphics{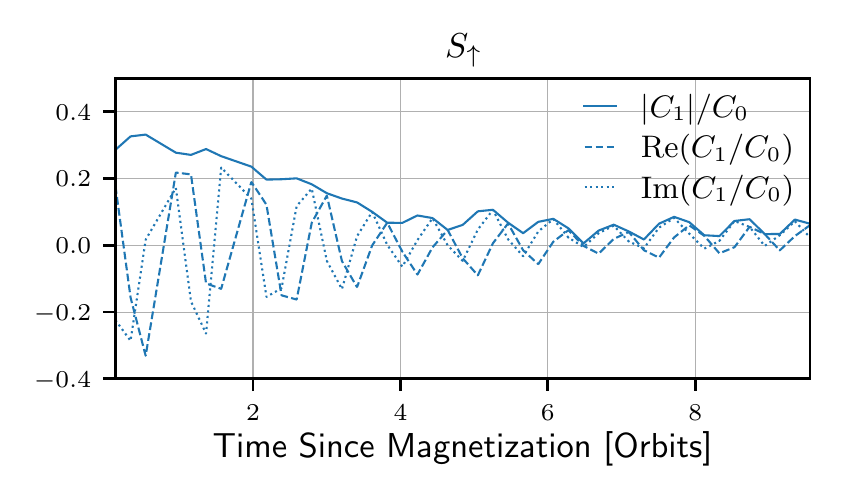}
\caption{Amplitude of the normalized $m=1$ density mode after magnetic field insertion for the spin-up case. Note that the components of the $m=1$ non-axisymmetric mode take on a slightly non-zero average value at late times in this case. This seems to be due to slight eccentricity that develops due to the non-axisymmetric variations in accretion that occur intermittently as a result of both the disk non-axisymmetry and the non-axisymmetry of the seed magnetic field. Our diagnostic measures just the angular $m=1$ mode, it does not account for radial distribution of matter and bears no relation to the disk center of mass moment. There is therefore no reason to expect it to always oscillate around zero, and this detail can be ignored as irrelevant to the dynamics of interest.}\label{fig:dens_mode_up}
\end{figure}
\begin{figure}[!htb]
\includegraphics{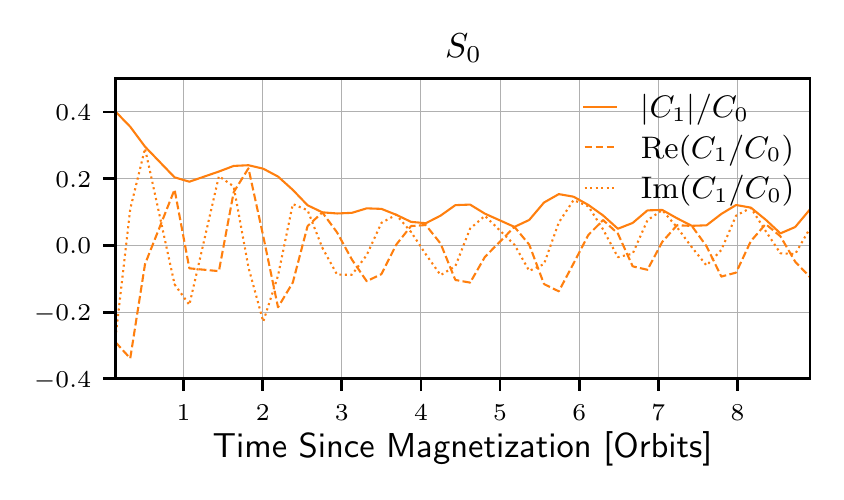}
\caption{Amplitude of the normalized $m=1$ density mode after magnetic field insertion for the non-spinning case.}\label{fig:dens_mode_0}
\end{figure}
As shown in Figure~\ref{fig:dens_mode_up} and Figure~\ref{fig:dens_mode_0}, the introduction of the seed magnetic field significantly reduces the amplitude of the dominant $m=1$ mode within the first four orbits.
After this decline, the $m=1$ mode amplitude of the magnetized cases has an average amplitude of roughly $\sim 1/5^{\mathrm{th}}$ their original value.
Our prior results for non-magnetized disks show that the saturation amplitudes of $m=1$ modes remain constant after saturation, and therefore at late times (specifically, at times more than 4 orbits after field insertion) the magnetized disks have lost $\sim 1/5^{\mathrm{th}}$ their amplitudes compared to the non-magnetized cases.
Table~\ref{tab:amp_compare} lists the exact average amplitudes compared to their initial values for both cases.

\renewcommand{\tabcolsep}{6pt}
\begin{table}[!hbt]
\caption{Average amplitudes of the $m=1$ mode ($\langle C_{1}/C_{0} \rangle_{\mathrm{late}}$) compared to their values shortly after PPI saturation and MRI seed-field insertion (Initial $C_{1}/C_{0}$). The last column shows the ratio between the late time average for the magnetized disks and the initial values. Late times are those 4 or more orbits after magnetic field insertion in both cases.}\label{tab:amp_compare} \def\arraystretch{1.5}
\begin{tabular}{cccc}
  \hline\hline
Model \ & Initial \bfseries{$C_{1}/C_{0}$} & \bfseries{$\langle C_{1}/C_{0} \rangle_{\mathrm{late}}$} & Ratio \\ 
\hline \SuM{}\ & 0.29 & $5.9 \times 10^{-2}$ & 0.21 \\
 \SoM{}\ & 0.40 & $9.3 \times 10^{-2}$ & 0.23 \\
 \hline\hline
\end{tabular}
\end{table}

\subsection{Comparison to prior fixed-spacetime study\label{section:comparison}}
To make closer contact with the fixed-spacetime (Cowling approximation) results of~\cite{bugli_papaloizoupringle_2018}, we use the normalized \textit{mode power} diagnostic (Eq.~\ref{eq:mode_power}).

A key point of comparison is Figure~11 of~\cite{bugli_papaloizoupringle_2018}, which shows the $m=1$ mode power over time for a purely hydrodynamic disk and a magnetized disk with the $m=1$ mode perturbed (jump-starting its early growth), as well as initially axisymmetric magnetized disks.
Both the hydrodynamic disks and the perturbed magnetized disks reach $P_{1}/P_{0} \approx 4-7 \times 10^{-2}$, but for the latter this is only a transient feature, and the $m=1$ power quickly settles to match that of the initially axisymmetric magnetized disks.
This is one of the most consequential findings of~\cite{bugli_papaloizoupringle_2018}: the presence of weak magnetic fields, through amplification by the MRI and subsequent effect on disk dynamics, does not merely \textit{suppress growth} of the PPI-unstable $m=1$ modes; they also quickly \textit{erase all traces} of any powerful $m=1$ modes that once existed.
By comparing these results to our own, we can begin to address one of the outstanding questions raised in the conclusion of~\cite{bugli_papaloizoupringle_2018}: whether this is also true for \textit{self-gravitating} disks.

\begin{figure}[!htb]
\includegraphics{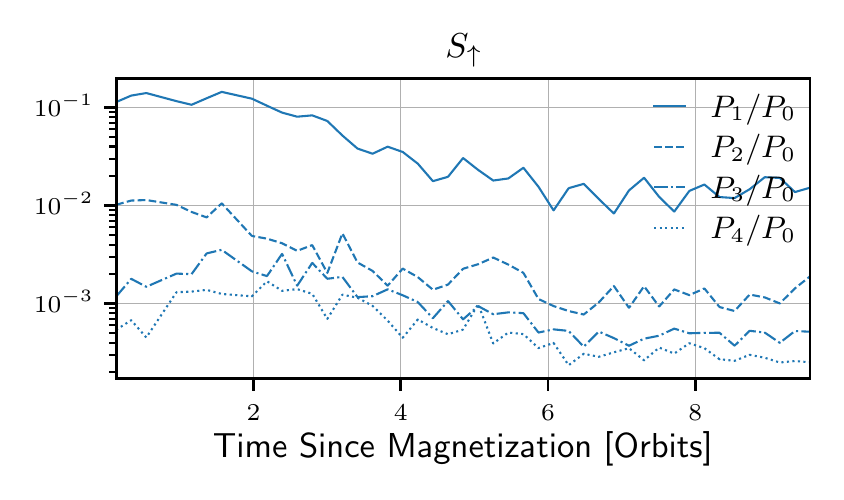}
\caption{The normalized density mode powers for $m=1, 2, 3, 4$ after magnetic field insertion for the spin-up case.}\label{fig:mode_power_up}
\end{figure}
\begin{figure}[!htb]
\includegraphics{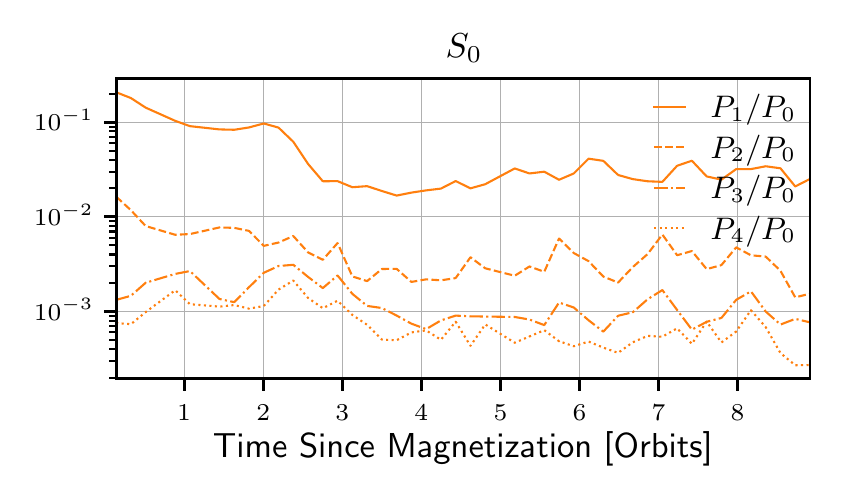}
\caption{The normalized density mode powers for $m=1, 2, 3, 4$ after magnetic field insertion for the non-spinning case.}\label{fig:mode_power_0}
\end{figure}

Figures~\ref{fig:mode_power_up},~\ref{fig:mode_power_0} show the mode powers for the first four modes for each of our disks.
Our self-gravitating disks share the preference for lower-$m$ modes seen in e.g. Figure~6 of~\cite{bugli_papaloizoupringle_2018}.
We start out with slightly stronger $P_{1}/P_{0} \approx 2 \times 10^{-1}$, and yet within a few orbits $P_{1}/P_{0}$ has been reduced by nearly an order of magnitude in both our evolutions.
Self-gravity and BH-disk interactions in our dynamic spacetime simulations do not seem to have protected the $m=1$ modes from being weakened, nor have they resulted in any significant increases to the $m=2, 3, 4$ modes that might preserve GW detectability.

\begin{figure}[!htb]
\includegraphics{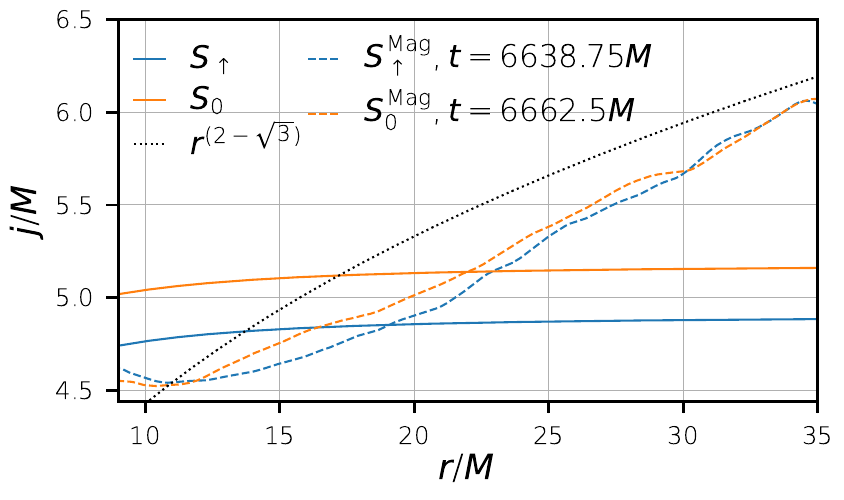}
\caption{Comparison of the equatorial radial profile of the specific angular momentum ($j$) between the initial configurations (solid lines) and magnetized configurations at late times (dashed lines). The black dotted line shows the steepest specific angular momentum profile that is PPI-unstable, configurations with steeper profiles will be stable to the PPI.}\label{fig:specific_j_compare}
\end{figure}

Are all traces of the initially powerful $m=1$ modes
\textit{completly} erased?  This question is more challenging to
answer completly in the absence of a resolution study. Our simulations
show a persistent $m=1$ mode in the MRI-saturated steady-state of our
disks, and this mode has about 10 times the power of those seen at
late times in the magnetized disks
of~\cite{bugli_papaloizoupringle_2018}, and is much more coherent as
well. However, while our azimuthal resolution is roughly comparable
to that of~\cite{bugli_papaloizoupringle_2018} in the region of the
disk, our vertical and radial resolution is just $\sim
1/3^{\mathrm{rd}}$ of theirs, notwithstanding the differences in
convergence between the different numerical methods.  These
differences could be important in light of the well known difficulties
of adequately resolving steady-state MRI
turbulence~\cite{hawley_assessing_2011,shiokawa_global_2012,hawley_testing_2013,blaes_general_2014,ryan_resolution_2017}. To understand what might be supporting the $m=1$ mode at late times,
we can look at the specific angular momentum profile. Figure~\ref{fig:specific_j_compare} shows that the profile has
steepened substantially toward the PPI-stable limit over the evolution, however in the
region within about $20 M$ where most of the disk material is the disks remain marginally PPI-unstable,
suggesting that the PPI might be responsible for the persistance of these modes.

Therefore, it is not straightforward for us to conclude with certainty
whether the relativly strong $m=1$ mode we observe even with magnetic
fields is attributable to self-gravity, and/or other differences with
the disk models used in~\cite{bugli_papaloizoupringle_2018}. Yet
another important difference between our evolutions and those
of~\cite{bugli_papaloizoupringle_2018} is that our disks are seeded
with a \textit{poloidal} magnetic field configuration while
in~\cite{bugli_papaloizoupringle_2018} disks with initally
\textit{toroidal} fields are evolved.

What is clear is that self-gravity is not a panacea for the ills the
MRI causes to PPI-supported $m=1$ modes: even disks with significant
self-gravity, starting from a highly asymmetric, $m=1$ dominated
state, can be quickly smoothed out by the presence of a weak poloidal
magnetic field.

\subsection{Gravitational wave signal\label{section:gw_signal}}

\begin{figure}[!htb]
\includegraphics{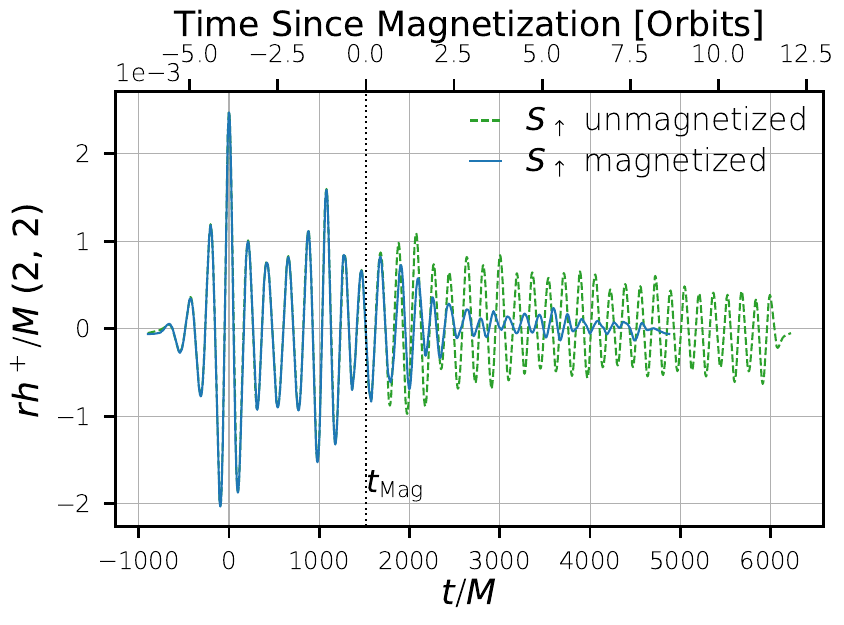}
\caption{Strain for the spinning BH case. Solid lines are the magnetized case, dashed lines are the non-magnetized case. Top time axis shows orbits since magnitization, bottom uses geometrized units and measures time from peak signal amplitude.}\label{fig:strain_sup}
\end{figure}

\begin{figure}[!htb]
\includegraphics{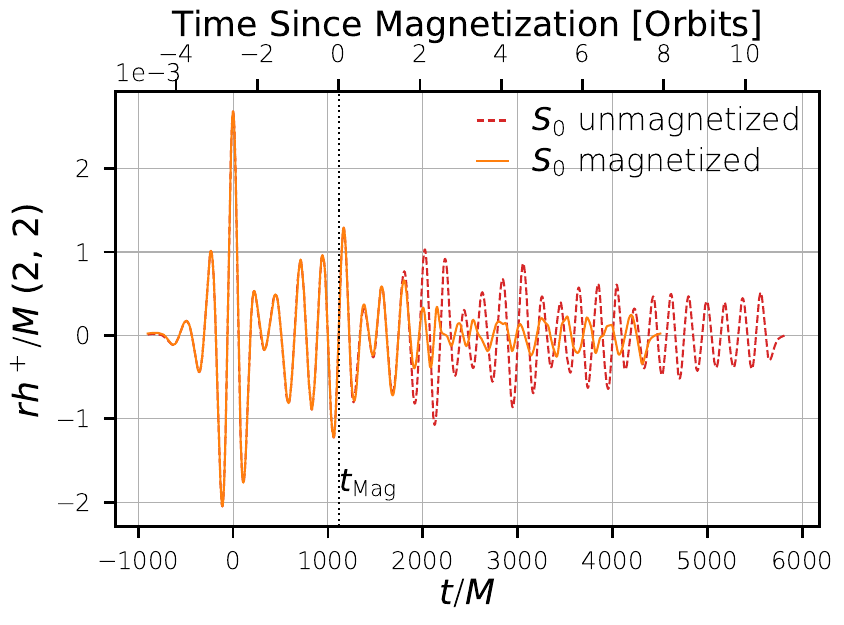}
\caption{Strain for the non-spinning BH case. Solid lines are the magnetized case, dashed lines are the non-magnetized case. Top time axis shows orbits since magnitization, bottom uses geometrized units and measures time from peak signal amplitude.}\label{fig:strain_s0}
\end{figure}

Figures~\ref{fig:strain_sup},\ref{fig:strain_s0} show the strain of
the (2, 2) mode extracted for each configuration.  The solid lines
indicate the case where magnetic fields are inserted, the dashed lines
show the strain from the original, non-magnetized evolutions.  In both
cases it can be seen that the reduced $m=1$ amplitude caused
by the magnetization of the disks translates into a reduced GW
amplitude.

\begin{figure}[!htb]
\includegraphics{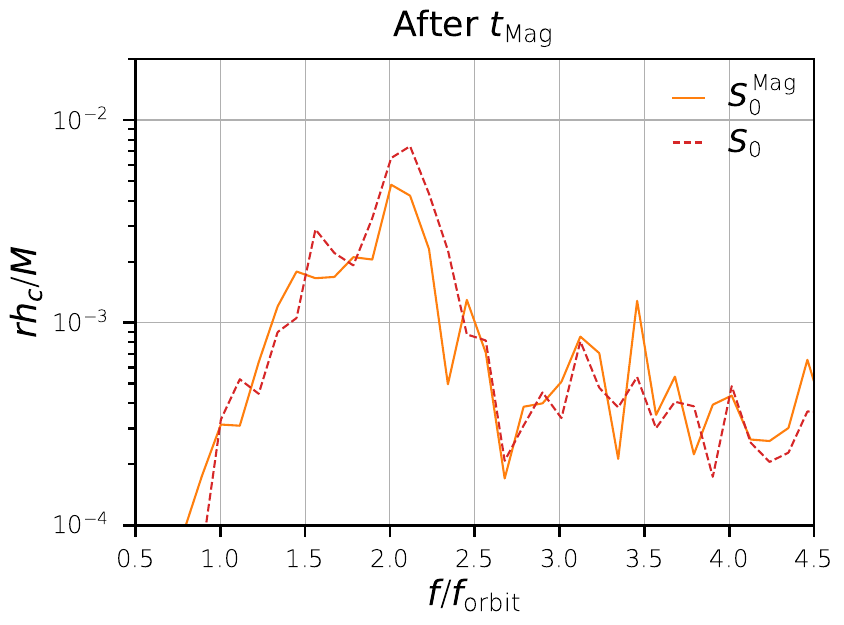}
\caption{Characteristic strain for the non-spinning BH case, for the GW signal emitted after the magnetic field insertion. Solid lines are the magnetized case, dashed shows the signal for the same times in the non-magnetized case.}\label{fig:char_strain_s0}
\end{figure}

\begin{figure}[!htb]
\includegraphics{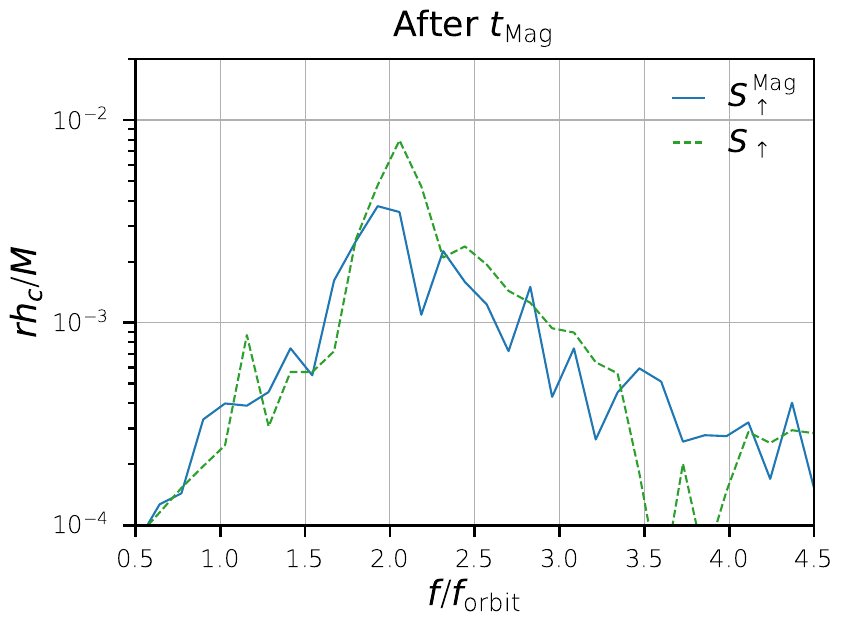}
\caption{Characteristic strain for the spinning BH case, for the GW signal emitted after the magnetic field insertion. Solid lines are the magnetized case, dashed shows the signal for the same times in the non-magnetized case.}\label{fig:char_strain_sup}
\end{figure}

To gain better insight to the effect this will have on detection, we look instead at the frequency-domain quantity of characteristic strain. This quantity is useful for assesing detectability, as the area under a characteristic strain curve and above a given noise floor roughly corresponds to the signal-to-noise ratio (SNR) for the given optimal matched filtering (see \cite{moore_gravitational-wave_2014} for details).
Characteristic strain is defined for positive frequencies only and can be computed as,
\begin{equation}\label{eq:char_strain}
  h_{c} = 2f|\tilde{h}_{\mathrm{res}}|,
\end{equation}
where $\tilde{h}_{\mathrm{res}}$ is the fourier transform of the interferometer response. Here we will assume the response to be,
\begin{equation}\label{eq:response}
  \tilde{h}_{\mathrm{res}} = \sqrt{\frac{|\tilde{h}_{+}|^{2} + |\tilde{h}_{\times}|^{2}}{2}}.
\end{equation}
This averages the detector response over polarizations, which is appropriate for assesing average SNR of sources whose arrival time and sky location are unknown.
We will therefore take $h_{c}$ computed from (\ref{eq:response}) and (\ref{eq:char_strain}) to be the \textit{polarization-averaged characteristic strain}.
Figures~\ref{fig:char_strain_sup},\ref{fig:char_strain_s0} show this quantity for the portion of the signals after the magnetic field insertion.
In both cases, the peak amplitude of the signal is reduced by a factor of about 2 from the non-magnetized case (which will generally imply a similar reduction to expected SNR).
While the amplitude is reduced, the frequency of the $m=1$ modes remains roughly double the orbital frequency, strongly suggesting that the PPI is still responsible for the maintenance of these modes, rather than them being the result of unrelated random fluctuations and turbulence in the disk.

\begin{figure}[!htb]
\includegraphics{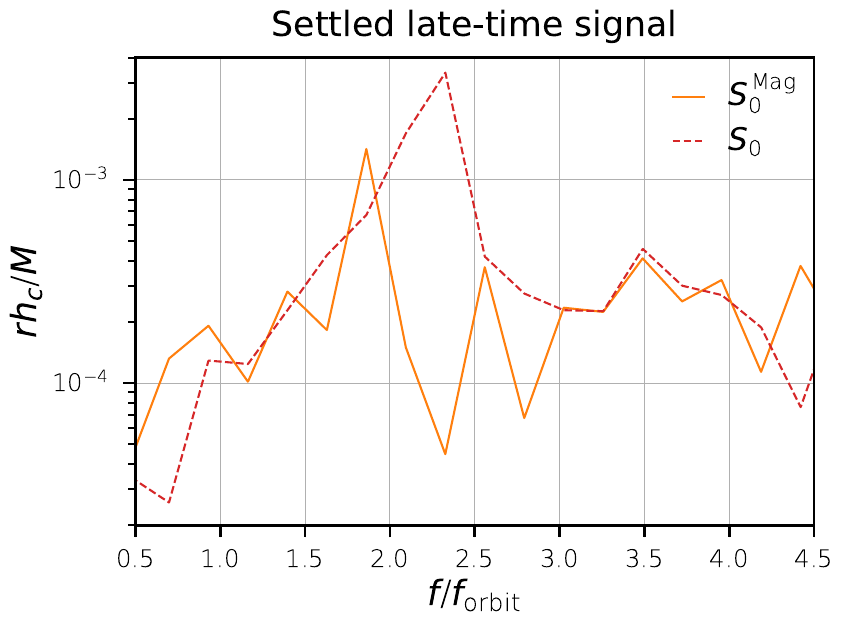}
\caption{Characteristic strain for the non-spinning BH case, for the GW signal emitted during the period when the disk has entered a settled steady-state. Solid lines are the magnetized case, dashed shows the signal for an equivalent period in the non-magnetized case.}\label{fig:char_strain_settled_s0}
\end{figure}

\begin{figure}[!htb]
\includegraphics{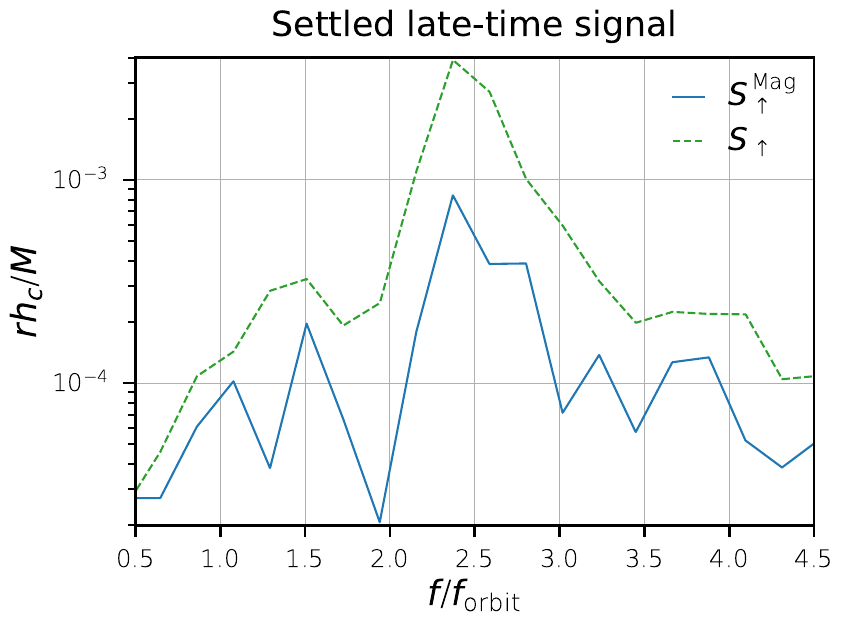}
\caption{Characteristic strain for the spinning BH case, for the GW signal emitted during the period when the disk has entered a settled steady-state. Solid lines are the magnetized case, dashed shows the signal for an equivalent period in the non-magnetized case.}\label{fig:char_strain_settled_sup}
\end{figure}

The time-series plots show that the loudest portions of the gravitational wave signals for the magnetized disks occur early on, as the amplitude of the $m=1$ mode is still in the process of being reduced.
This early time is a period of transition, and transient dynamics tend to depend sensitively on the inital conditions.
Since we mean for our disks to act as stand-ins for the potential end-states of a wide range astrophysical events, it is wise to avoid paying too much attention to the transient dynamics, and instead focus on the late time steady-state, which we can hope to be more generic.
We define the settled, late-time part of the signal to start $2500M$ (roughly 6 orbits) after the maximum GW amplitude for the magnetized disks. The magnetized disks have their seed fields inserted shortly after PPI saturation, so this cutoff time also ensures that the signal starts nearly two orbits after MRI saturation and doesn’t capture any transients of MRI growth in addition to avoiding transients of PPI saturation. These signals are obviously shorter than the full duration GW signals, with durations of $2000M$ and $2500M$ for \SoM~and \SuM, respectively, so they are compared with the last $2000M$ of \So~and the last $2500M$ of \Su, to ensure a meaningful comparison of the polarization-averaged characteristic strains in Figures~\ref{fig:char_strain_settled_sup},\ref{fig:char_strain_settled_s0}.
As is apparent from these figures, the effect of magnetization is even more extreme for this part of the GW signal, with the amplitudes reduced by factors closer to \(\sim 4-5\).

\subsection{Detectability estimates\label{section:detect}}
\begin{figure*}[!tb]
  \includegraphics{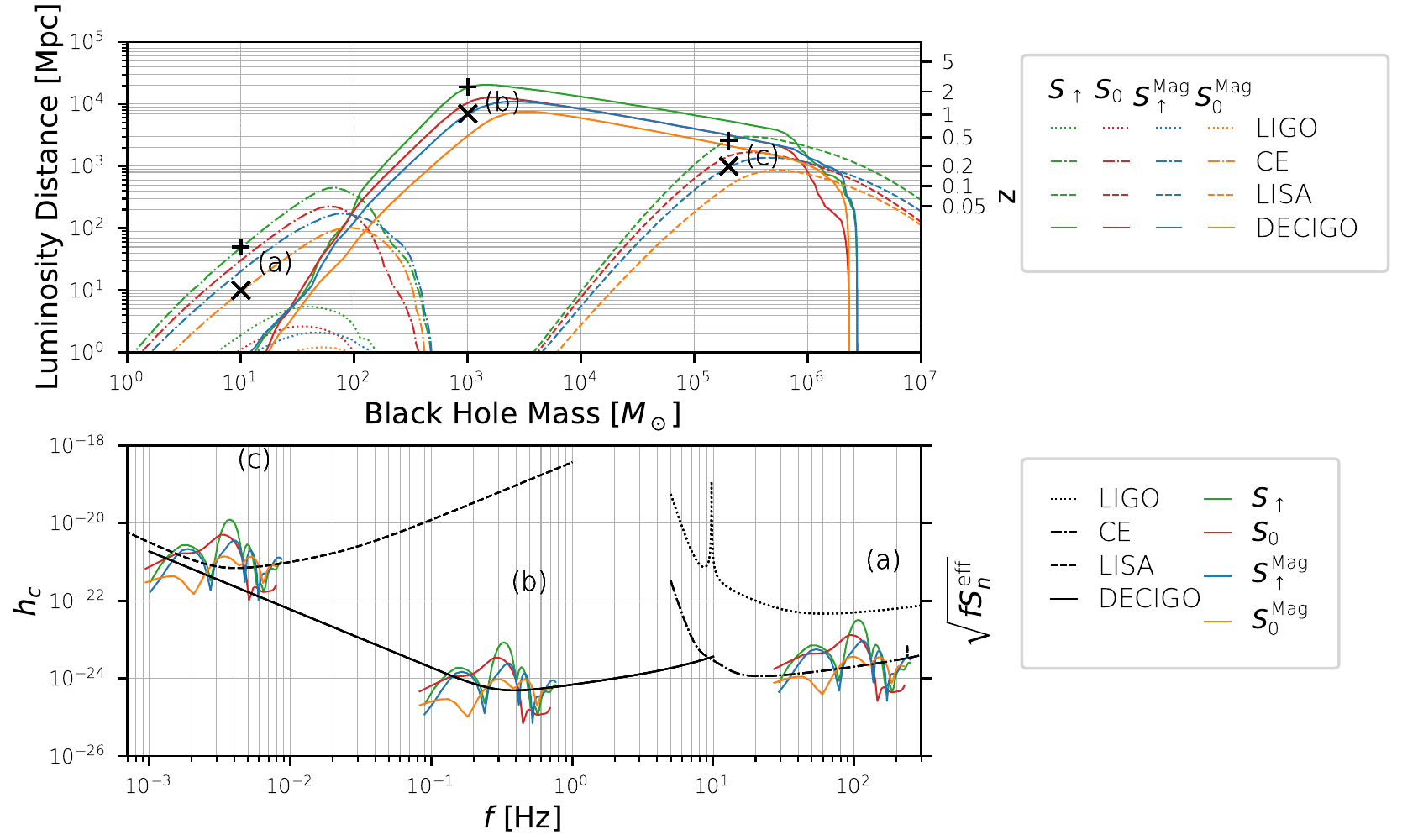}
  \caption{Comparison of the detectability by GW observatories LIGO, CE, DECIGO, and LISA, for optimally-oriented BH-disks that are purely hydrodynamic (\Su, \So) versus those that were seeded with magnetic fields (\SuM, \SoM), for spinning and non-spinning central BHs respectively. Labels indicate features corresponding to three mass scales of astrophysical interest: (a) $10 M_{\odot}$, (b) $1000 M_{\odot}$, (c) $2 \times 10^{5} M_{\odot}$. Top: horizons of maximum detectable distance for each system mass for each observatory. The threshold for a system at a given mass and distance on the chart being considered detectable is that its GW signal achieves a sky and polarization averaged SNR $\geq8$ (given an optimal matched-filter). Line styles denote observatories, while colors indicate which of the four BH-disk configurations studied the GW signal originates from. The $+$ ($\times$) labels show the furthest detectable distance of \textit{non}-magnetized (\textit{magnetized}) configurations for each of the labeled mass scales: (a) $50$Mpc ($10$Mpc) by CE, (b) $19000$Mpc ($7000$Mpc) by DECIGO, (c) $2600$Mpc ($1000$Mpc) by LISA. Bottom: Characteristic strain curves for signals at the labeled mass-scales and distances marked by the $+$ symbols on the top plot. These strain curves are overlaid on top of noise curves for each of the detectors, and the area between curves is a proxy for the SNR.}\label{fig:det_horizon_sim_only}
\end{figure*}
Finally, we assess what the results of our numerical evolutions mean for the prospects of detecting gravitational waves from real astrophysical disks with similar configurations.
To do this we take the GW signals from the configurations we simulated, scale them to a range of different masses, and then compute their amplitude and red-shift for a range of source distances in order to estimate the signal-to-noise ratio (SNR) that can be expected from next-generation GW observatories.
Because our simulations do not extend for the full lifetime of the disks, we also construct a model that extrapolates the signal to its estimated full duration, and perform the same analysis with the extended waveforms.
We did this previously in~\cite{wessel_gravitational_2021} for the non-magnetized disks, here we follow the exact same proceedure for both magnetized and non-magnetized disks in order to determine the impact magnetization might have on detectability.

Given that we do not perform a resolution study, we regard the
estimates provided here as a \textit{rough upper bound} on
detectability for real BH-disk systems of comparable configurations,
in accordance with the discussion in Section~\ref{section:resolution}.

\subsubsection{Computing rescaled signals at detector\label{section:detector_frame}}
Because we modeled disk material as a perfect fluid with a
$\Gamma$-law EOS (Section~\ref{section:non-mag_ev}), the
magnetohydrodynamic and GR equations are invariant under uniform
rescaling of mass, which serves as the unit of length and time in
geometrized units. Thus, by simply adjusting the mass-scales of the relevant
observables, we are able to use each of our simulations to represent an entire family of BH-disk systems
that share the same dimensionless properties of BH spin magnitude, orientation, and BH-to-disk mass ratio.
For GWs, the strain
amplitude is proportional to system mass, while the frequency is
inversely-proportional, so applying these re-scalings to our GW model
for $rh(t/M, \theta, \phi)$ is straightforward. To go from $rh(t, \theta, \phi)$ in the source frame to the observed
strain at a detector, we first choose a value $r$ that will be the luminosity distance to the source
in the detector frame. Then the ratio of $M/r$ controls the overall amplitude of the strain.

For example, if we chose $M = 10 M_{\odot}$ and $r = 300\mathrm{Mpc}$, then we would plug $M [\mathrm{s}] = 10 M_{\odot}[\mathrm{kg}] \cdot \frac{G}{c^{3}}$, and $M [\mathrm{Mpc}] = 10 M_{\odot}[\mathrm{kg}] \cdot \frac{G}{c^{2}}\cdot\frac{\mathrm{Mpc}}{3\times10^{22}\mathrm{m}}$, into the below formula to compute $h$,
\begin{equation}\label{eq:h_without_redshift}
  h(t[\mathrm{s}], \theta, \phi) = rh(t[\mathrm{s}] / M[\mathrm{s}], \theta, \phi) \frac{M [\mathrm{Mpc}]}{300 \mathrm{Mpc}}.
\end{equation}

The above procedure accounts for the mass-scale and luminosity distance only.
To accomadate large cosmological distances, we also
shift the observed frequencies by an amount determined by the redshift
$z$, which we calculate from the luminosity distance in a standard
$\Lambda$CDM cosmology with $H_{0} = 67.7
\mathrm{km}/\mathrm{s}/\mathrm{Mpc}$ and $\Omega_{m} = 0.3089,
\Omega_{\Lambda} = 0.6911$.  The dominant mode of the GW signal is the
$(2, \pm2)$ mode, so we will only consider it in this analysis. To get
a sense of average detectability, we compute the strain for observers
viewing the source from an angle of $\theta = \pi/2.34$ relative to
the disk orbital plane, for which the amplitude of the (2,2) mode
equals its $\theta$-averaged value.  From this we compute the
polarization-averaged response at the detector
$\tilde{h}_{\mathrm{res}}$ via (\ref{eq:response}), and then the
characteristic strain $h_{c}$ via (\ref{eq:char_strain}), for which we
may then compute the sky-averaged SNR (assuming an optimal matched
filter) given detector parameters for Advanced
LIGO~\cite{prospects_2020}, Cosmic Explorer~\cite{reitze_cosmic_2019},
DECIGO~\cite{sato_status_2017}, and
LISA~\cite{amaro-seoane_laser_2017}.  The sky-averaged sensitivity
factor for a $90^\circ$ interferometer (equation 51
of~\cite{moore_gravitational-wave_2014}) along with the sensitivity
curves provided by~\cite{ligo_and_CE_sensitivity_data} were used for
LIGO and CE, while the approximate analytic sky-averaged
sensitivities\footnote{When comparing ground and space-based
detectors, a clash of conventions is encountered. For ground-based
detectors, averaging over polarization is assumed to be part of the
definition of characteristic strain $h_{c}$; for space-based
detectors, it is included as part of the sky-averaged response. Here
we compromise by multiplying the space-based sensitivities by a factor
of $\sqrt{2}$ to account for the $1/\sqrt{2}$ in $h_{c}$, effectively
removing the contribution of polarization averaging. This rescues the
relationship between SNR and the area between the $h_{c}$ and
sensitivity curves.} in~\cite{yagi_detector_2011}
and~\cite{robson_construction_2019} were used for DECIGO and LISA,
respectively.

\begin{figure*}[!htb]
  \includegraphics{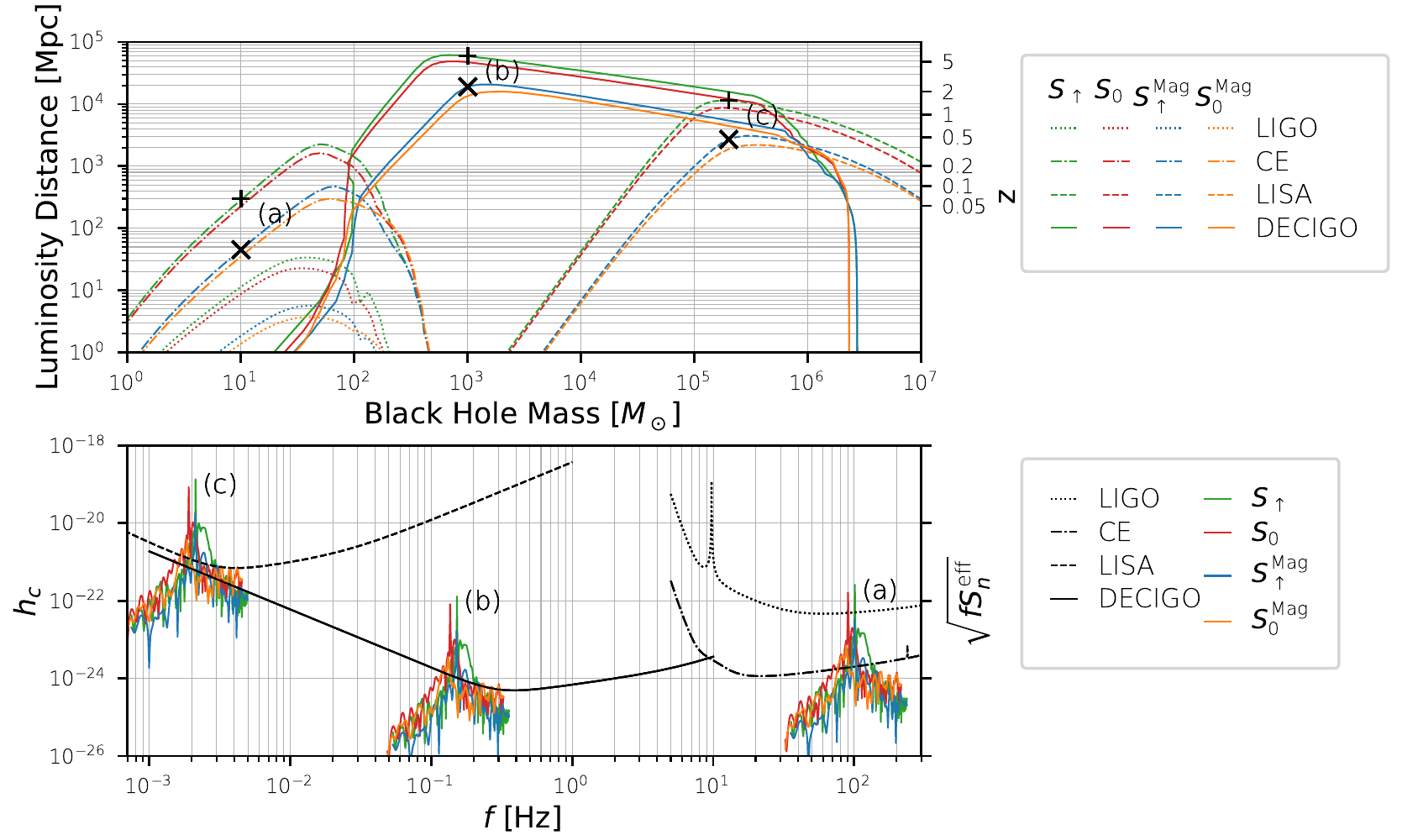}
  \caption{The same quantities as in Figure~\ref{fig:det_horizon_sim_only}, but this time for the model waveforms produced by extrapolating our simulated GWs to their expected full duration. Three mass-scales of astrophysical interest are again labeled: (a) $10 M_{\odot}$ (b) $1000 M_{\odot}$ (c) $2 \times 10^{5} M_{\odot}$. In the top panel, sources with these masses are marked by $+$ ($\times$) to denote the maximum distances \textit{non}-magnetized (\textit{magnetized}) BH-disk configurations of each mass could be detected: (a) $300$Mpc ($45$Mpc) by CE, (b) $60,000$Mpc ($19,000$Mpc) by DECIGO, (c) $11,500$Mpc ($2700$Mpc) by LISA.}\label{fig:det_horizon_full}
\end{figure*}

\subsubsection{Masses and distances of potential sources}\label{section:mass_scales}
We motivate the choice of mass scale in our detectability analysis by
considering three potential formation channels for BH-disk systems
similar to those we simulated.  At the low-end, BH-disk systems of
3-20 $M_{\odot}$ could result following NSNS and BHNS
mergers~\cite{voss_galactic_2003}, and are expected to be reasonably
common in the nearby universe.  Systems massing 25-140 $M_{\odot}$ and
250 $M_{\odot}$ or greater could be the end-results of Population III
stars ending their lives as collapsars
\cite{macfadyen_collapsars_1999,macfadyen_supernovae_2001,heger_nucleosynthetic_2002,heger_how_2003}.
These stars are expected to have peak formation rates in the $z \sim 5
- 8$ range~\cite{tornatore_population_2007,johnson_first_2013}, and
there is observational evidence for them at $z \approx 6$
\cite{sobral_evidence_2015}.  The highest-mass systems of $10^{3}
M_{\odot}$--$10^{6} M_{\odot}$ could concievably result from the
collapse of supermassive stars (SMS), simulations of which have been
shown to produce BHs surrounded by disks of about 10\% the BH
mass~\cite{shapiro_collapse_2002,shibata_collapse_2002,shapiro_collapse_2004,sun_magnetorotational_2017,uchida_gravitational_2017}.
These events have been proposed to occur in the early universe and
provide the seeds for the formation of supermassive BHs that emerge at
$z \approx 7$
\cite{loeb_collapse_1994,shapiro_relativistic_2003,koushiappas_massive_2004,shapiro_spin_2005,begelman_formation_2006,lodato_supermassive_2006,begelman_evolution_2010}
(see reviews
\cite{haiman_formation_2013,latif_formation_2016,smith_first_2017}).

In addition to considering a range of masses that encompases all the scales listed above, we also choose a set of three distinct masses to act as representatives for each of the three channels above.
These scales, as labeled in Figure~\ref{fig:det_horizon_sim_only} and Figure~\ref{fig:det_horizon_full} in the subsequent subsections, are:
\begin{enumerate}[label=(\alph*)]
  \item $M = 10 M_{\odot}$, representing the possible aftermath of a BHNS merger.
  \item $M = 1000 M_{\odot}$, representing the possible aftermath of a
    low-mass SMS collapse.
  \item $M = 2 \times 10^{5} M_{\odot}$, representing the possible aftermath of an high-mass SMS collapse.
\end{enumerate}


\subsubsection{Detectability horizons of settled GW signals}\label{section:settled_signal}

Since we aim to gain insight into the GW detectability of a wide class
of phenomena that may lead to non-axisymmetric disks, including the
early time transient GW signals is inappropriate.  Therefore, we only
assess the detectability of the \textit{settled} late-time dynamics,
defined as the portion of the signal $2500 M$ after the maximum GW
amplitude for the magnetized cases, as in
Section~\ref{section:gw_signal}.  To enable meaningful comparison
between all signals, here they are all truncated to the same end time,
which is set to be $4500 M$ by our shortest simulation, and therefore
all the settled late-time signals in this section have a $2000 M$
duration. This is notably shorter than the duration of the signals
analyzed in~\cite{wessel_gravitational_2021}, and results in lower
maximum detection distances. However, since our simulations do not
capture the full signal duration anyway, the absolute detectability
distances are less important than relative differences between the
configurations.

In Figure~\ref{fig:det_horizon_sim_only} we show the maximum
detectability distances for these signals accross a range of masses
and propagated from a range of distances. That is: for each simulated BH-disk system we choose a set of mass-scales, and for each of those masses we compute the GW strain in the detector frame for a range of luminosity distances (following the methods of Section~\ref{section:detector_frame}). We then use the detector-frame strain to compute the SNR for each observatory, assuming an optimal matched
filter. The lines in the top panel of the figure show the maximum distances systems at each mass were found to be detectable, where we have chosen an SNR of 8 as the
threshold of detectability for this analysis.
We see that the spinning cases \Su\ and \SuM\ are more
detectable than their non-spinning counterparts \So\ and \SoM.
In~\cite{wessel_gravitational_2021} we showed this was accounted for
by the increased orbital frequency of the disk in the \Su\ case, due to
the smaller ISCO radius of the co-spinning BH, so it is not surprising
that this carries over to \SuM.  The difference between the
non-magnetized and magnetized configurations is stark: accross all
masses the detectability horizon has been reduced by factors of $\sim
3$--$5$ for the magnetized disks.  While the accretion rate is
somewhat larger in the magnetized cases, for the times represented
here little accretion has had time to occur and the disks are nearly
the same masses.  The difference in detectability is therefore
predominantly due to the suppression of the $m=1$ modes that occurs
during the growth of the MRI in the magnetized disks.

\subsubsection{Detectability horizons of extrapolated GW models\label{section:signal_model}}
\begin{figure}[!tb]
  \includegraphics{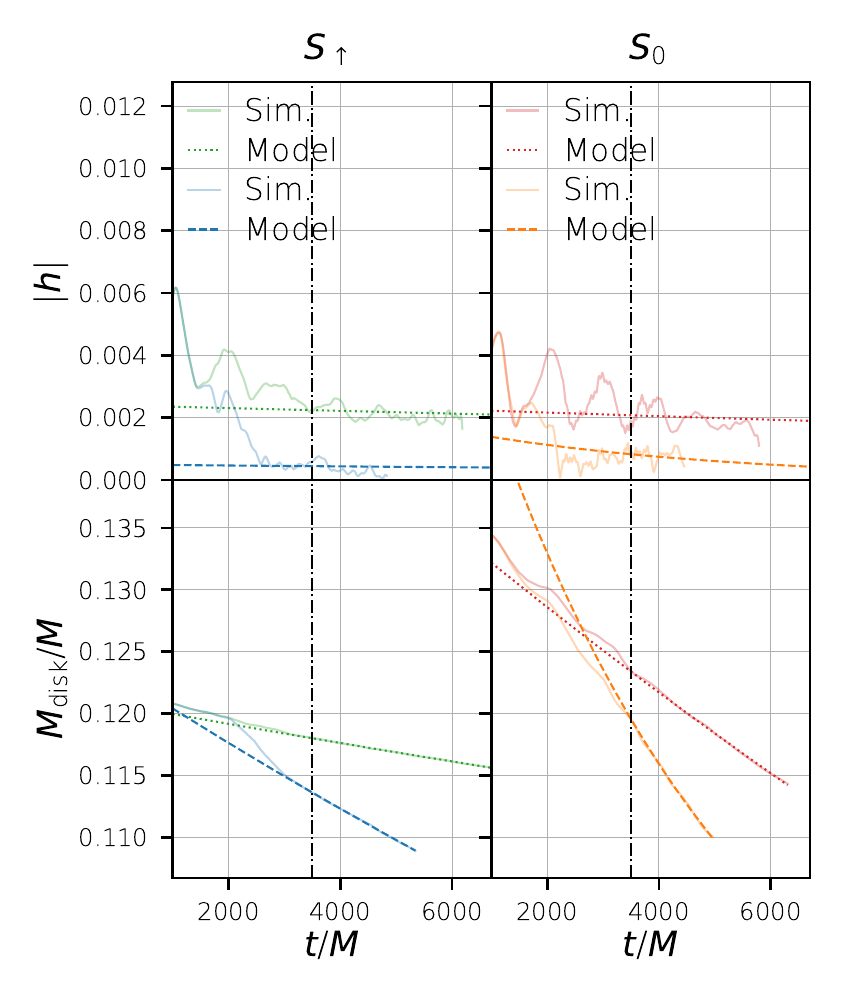}
  \caption{Comparison of analytic model to quantities extracted from simulations for both magnetized and non-magnetized accretion disks (indicated by line colors). Dashed curves show analytic model fit to the magnetized disks, dotted curves show model fit to the non-magnetized disks, and solid lines show the quantities being fit: strain amplitude in the top row, disk rest-mass in the bottom row. Analytic models are fit using only data after the time marked by the vertical dash-dotted lines, to avoid influence of early-time transients on the extrapolated late-time signal model.}\label{fig:model_fit}
\end{figure}

The actual detectability of GW signals depends on both their amplitude and duration.
Therefore, we need to extrapolate beyond the short duration we simulated to get a model of the full signal for each case.
To do this, we use the same simple signal model as~\cite{wessel_gravitational_2021}: motivated by the quadrapole formula, the signal at late times is assumed to be of the form,
\begin{equation}\label{eq:gw_model}
  rh = B \exp((i \omega_{0} - \gamma) t + i \phi_{0}).
\end{equation}
The amplitude, $B$, of the waveform is set by the amplitude of the extracted GWs, but the exponent of amplitude falloff, $\gamma$, is extracted by fitting the disk mass at late times to a decaying exponential.
The frequency $\omega_{0}$ and initial phase $\phi_0$ are set by matching to the unrolled phase profile of the GW signal, and extrapolating it past the final time with a linear best-fit.
The model fitting is shown in Figure~\ref{fig:model_fit}.
We cannot know for certain how long the GW-producing $m=1$ modes will persist after the end of our evolutions, so we conservatively estimate that the GW signal terminates when only 10\% of the original disk mass remains, once again using a smooth falloff to avoid artifacts from the imposed cutoff.

We now take the settled late-time GW signals (after the cutoff of $2500 M$), and instead of truncating them at $4500 M$ as we did in the previous section, we use their full duration.
In the last $500 M$, we smoothly match onto the model waveform fitted for that case, which then extends the signal past the end of our simulations.

We perform the same analysis as in the previous section to these extended model waveforms, and the results are shown in Figure~\ref{fig:det_horizon_full}.
Here, the differences in detectability are even more noticable: the maximum detection distances are reduced by factors of $\sim 3.1$--$6.6$ for the magnetized disks.
This can at least partially be attributed to the increased accretion rate caused by the magnetic shear, which dramatically shortens the lifetime of the disk extrapolated by the model.
However, although the change in signal lifetime is significant, its impact on detectability is not that large because the most of the GW signal power comes from the early-time emission, so the reduction of the $m=1$ amplitude is the dominant effect.

Now that the GW signals are not cut off prematurely by the end of the
simulation, the estimates of maximum detectability distances are more
realistic. In all cases the spinning BH configurations are most
detectable, and in Table~\ref{tab:max_det_dists} we list the estimated
maximum detection distances for each of the mass-scales of interest
introduced in Section~\ref{section:mass_scales}.

\renewcommand{\tabcolsep}{3pt}
\begin{table}[!tb]
\caption{Maximum detection luminosity distances for the \Su{} and \SuM{} BH-disk configurations for a set of masses motivated by plausible formation channels (see Section~\ref{section:mass_scales}). In each case, we consider only the detector that achieves its greatest detection distance for the given source. In all cases the singnal is not powerful enough for detection by Advanced LIGO, so it is not included here.}\label{tab:max_det_dists} \def\arraystretch{1.5}
\begin{tabular}{cccccc}
  \hline\hline
  \multirow{2}{*}{Label} & \multicolumn{1}{c}{BH Mass} & \multicolumn{3}{c}{Maximum Distance (Mpc)} & \multirow{2}{*}{Observatory} \\ \cline{3-5} & ($M_{\odot}$) & \Su{} & \SuM{} & $S_\uparrow / S^\mathrm{Mag}_\uparrow$ \\ 
\hline
  (a) & 10 & 300 & 45 & 6.67 & CE \\
  (b) & 1000 & 60000 & 19000 & 3.16 & DECIGO \\
  (c) & \(2\times10^{5}\) & 11500 & 2700 & 4.26 & LISA \\
 \hline\hline
\end{tabular}
\end{table}

\section{Discussion\label{section:discuss}}
In this study, we conducted magnetohydrodynamic simulations in
dynamical spacetime comparing the dynamics of self-gravitating
accretion tori around spinning and non-spinning BHs for both
magnetized and unmagnetized configurations. Our goal was to estimate a
rough upper bound on the resilience of the non-linear saturated state
of the PPI in self-gravitating disks to the effects of the MRI on the
disk dynamics, so MRI-susceptable field configurations were only
seeded once PPI modes reached their maximum amplitude, giving the PPI
its best chance of survival.  Despite this, these disks, which were
previously shown to be powerful sources of gravitational
radiation~\cite{kiuchi_gravitational_2011,wessel_gravitational_2021}
due to their persistent $m=1$ non-axisymmetries, reacted to the
introduction of magnetic fields by exhibiting quick MRI growth and
strong suppression of the $m=1$ mode and the associated gravitational
radiation.  Witnessing such strong and sudden suppression of these
modes, even in the fully non-linear self-gravitating saturation state,
suggests that suppression of the $m=1$ mode will generally occur for a
wide range of conceivable configurations with less developed PPI
modes, whenever weak magnetic fields of similar magnitude are present.

These results are similar to the findings
of~\cite{bugli_papaloizoupringle_2018} for disks without self-gravity
around non-spinning black holes.  By constructing a mode-power
diagnostic, we are able to approximately compare the power of the
low-$m$ non-axisymmetric modes in our disks to those reported
by~\cite{bugli_papaloizoupringle_2018}.  We find that the PPI
supported $m=1$ mode is quickly reduced in amplitude, dropping nearly
an order-of-magnitude in four orbits in both \SuM{} and \SoM{}
configurations.  Tracking the components of magnetic field flux, we
see fast increases in the azimuthal and radial components for the
first two orbits, characteristic of MRI
growth~\cite{balbus_instability_1998}.  After this, the fields
saturate, then decay.  It is notable (but not surprising) that the
majority of the $m=1$ amplitude reduction happens during the first few
orbits, when the magnetic fields are strongest. At late times, the
strength of the fields is determined by the equilibrium between
dissapation (purely numerical in our simulations) and MRI-driven
turbulence~\cite{balbus_instability_1998,blaes_general_2014}.  In our
simulations, the field strength peaks before it relaxes to weaker magntudes not high above those at initialization,
and this seems to allow the PPI to support the $m=1$ modes against
further disruption, resulting in $m=1$ modes of about 10 times the
power of those seen at late times
by~\cite{bugli_papaloizoupringle_2018}.

When interpreting this result, it is important to consider the known
difficulties in adequately resolving steady-state MRI
turbulence~\cite{hawley_assessing_2011,hawley_testing_2013,ryan_resolution_2017}.
As detailed in Section~\ref{section:comparison}, our simulations are
slightly lower in resolution in some directions than those
of~\cite{bugli_papaloizoupringle_2018}, so a careful resolution study
would need to be carried out for error estimates in our results.
However, the lack of a resolution study does not prevent the results
reported here from providing a correct qualitative description of the
ability of saturated $m=1$ PPI modes in massive, self-gravitating
disks to survive the transient growth of the MRI.  This is because
greater resolution is expected to make the average magnetic field
stronger, and the dynamical influence of the MRI and associated
turbulence should be even more
pronounced~\cite{hawley_testing_2013,ryan_resolution_2017}, while the
PPI is already well-resolved.

Our key finding is therefore that the MRI's ability to reduce the
amplitude of large non-axisymmetric features is robust even in the
non-linear self-gravitating case, and even when starting from extremly
non-axisymmetric disks.  Because high-amplitude orbiting
non-axisymmetric patterns are necessary to produce powerful GWs, this
result reduces our expectation that such systems will act as
detectable GW sources at the large distances determined by
unmagnetized models.

To assess detectability, we fit a simple GW model to the extracted
gravitational waveforms, and used our study of the observed disk
dynamics to extrapolate an optimistic case where the MRI reduces but
does not eliminate the disk's $m=1$ mode, allowing GW production at a
reduced amplitude, and for a duration shortened by the more rapid
accretion of the magnetized configurations.  This optimistic case
predicts reductions of the maximum luminosity distances of detection
by factors ranging from 3.16 to 6.67.  For non-magnetized disks, the
spinning BH configuration \Su{} was detectable by DECIGO out to
red-shifts of $z \approx 5$ for systems of mass $1000 M_{\odot}$,
however the most detectable magnetized configuration \SuM{} produces a
signal too weak to be detected by DECIGO beyond $z \approx 2$, a 76\%
reduction in luminosity distance.

Finally, we caution that our study does not rule out the potential of
accretion disks as engines of detectable gravitational waves.  We have
shown that a particular non-axysymmetric disk configuration, which
acts as a powerful GW source when not magnetized, quickly becomes
significantly more axisymmetric when the MRI is allowed to develop,
greatly reducing GW emission.  This does not prove that there are no
other magnetized disk configurations that could be sustained soruces
of GW production. The task of finding such configurations, or ruling
out their existence, remains to be done.

\begin{acknowledgments}
The following grants supported this work: NSF Grants PHY-1912619 and PHY-2145421, and NASA Grant 80NSSC22K1605 to the University of Arizona;
NSF Grant PHY-2006066 and NASA Grant 80NSSC17K0070 to the University of Illinois at Urbana-Champaign.
M.R. acknowledges also support by the Generalitat Valenciana Grant CIDEGENT/2021/046 and by the Spanish Agencia Estatal de
Investigaci\'on (Grant PID2021-125485NB-C21).
Extreme Science and Engineering Discovery Environment
(XSEDE) Grant TG-PHY190020 provided the high performance computing resources.
XSEDE itself is supported by NSF Grant ACI-1548562.
A.T. acknowledges support from the National Center for Supercomputing Applications (NCSA) at the University of Illinois at Urbana-Champaign through the NCSA Fellows program.
Simulations were run on the Stampede2 cluster, maintained and operated by the Texas Advanced Computing Center (TACC) at the University of Texas at Austin, supported by NSF Grant ACI-1540931.
Additional simulations and data analyses were run on the University of Arizona's (UArizona) Ocelote cluster, which is maintained by the UArizona Research Technologies department and supported by UArizona TRIF, UITS, and Research, Innovation, and Impact (RII).
\end{acknowledgments}

\appendix*
\section{Evolution of constraint violations\label{appendix:constraints}}

\begin{figure}[!tb]
  \includegraphics{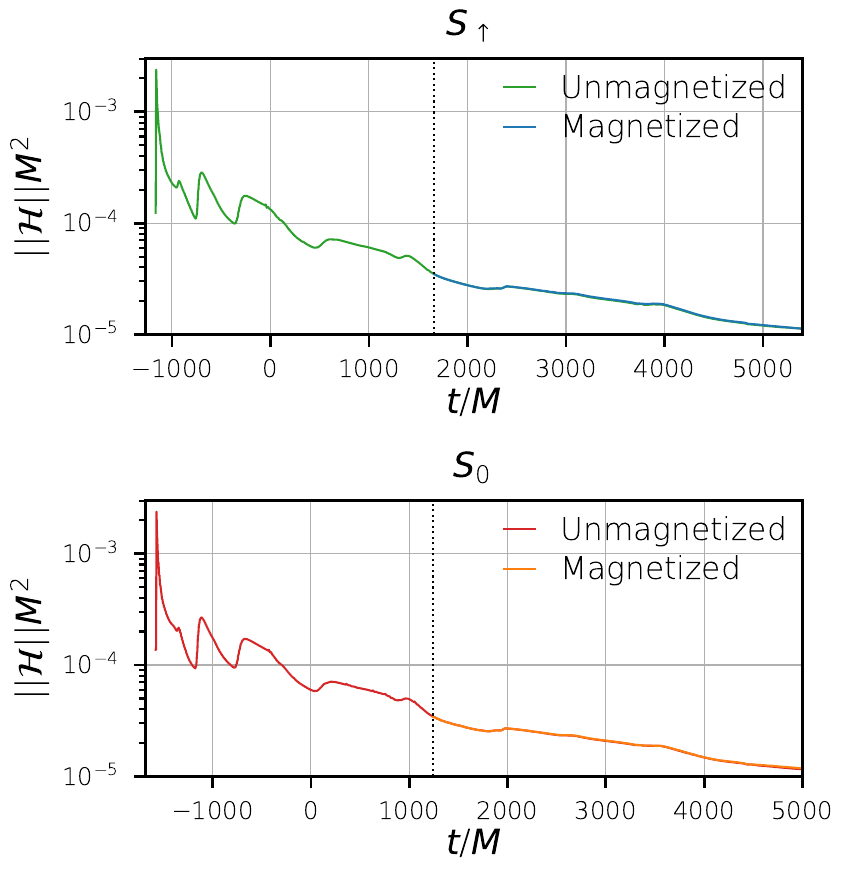}
  \caption{Evolution of the $L_{2}$ norm of the Hamiltonian constraint violations for both configurations, with the magnetized disk evolutions overlayed on the non-magnetized evolutions.}\label{fig:ham_con_comp}
\end{figure}

\begin{figure}[!tb]
  \includegraphics{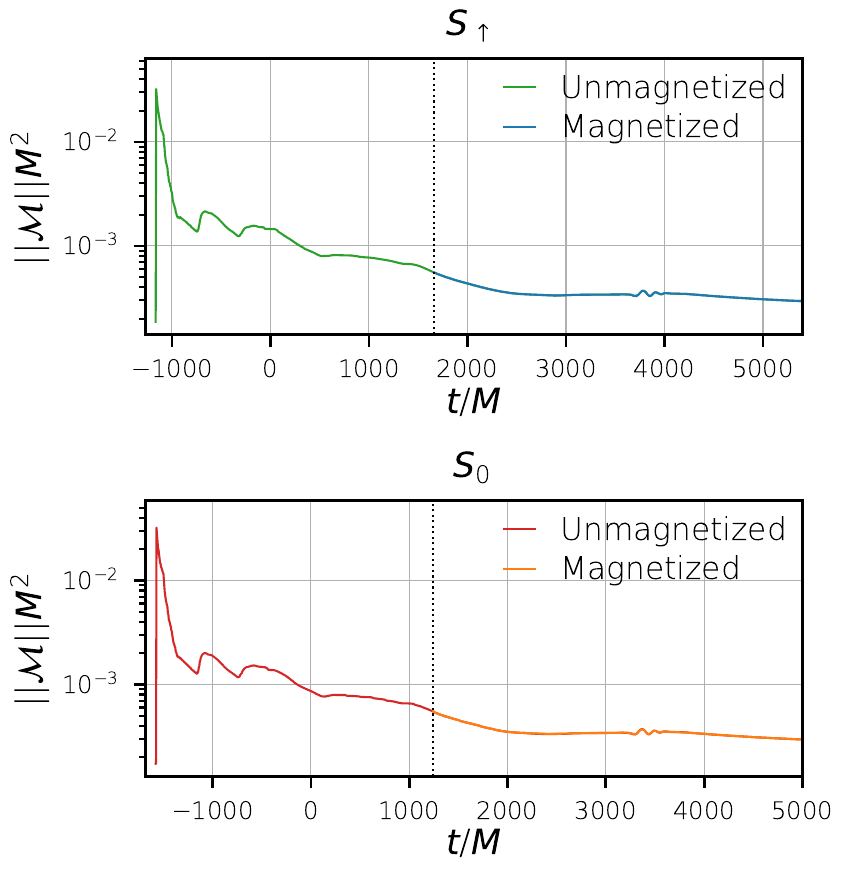}
  \caption{Evolution of the $L_{2}$ norm of the momentum constraint violations for both configurations, with the magnetized disk evolutions overlayed on the non-magnetized evolutions.}\label{fig:mom_con_comp}
\end{figure}

Throughout the evolutions we track the constraint violations to ensure the physical validity.
In terms of the BSSN variables, the Hamiltonian constraint is,
\begin{equation}\label{eq:ham_con}
  \mathcal{H} = \tilde{\gamma}^{ij}\tilde{D}_{i}\tilde{D}_{j}e^{\phi} - \frac{e^{\phi}}{8}\tilde{R} + \frac{e^{5\phi}}{8}\tilde{A}_{ij}\tilde{A}^{ij} - \frac{e^{5\phi}}{12}K^{2} + 2\pi e^{5\phi}\rho.
\end{equation}
We monitor the $L_{2}$ norm of $\mathcal{H}$, defined as the square root of the integral of $|\mathcal{H}|^{2}$ over the volume, which we here denote $||\mathcal{H}||$.
Figure~\ref{fig:ham_con_comp} shows the values of this constraint for the purely hydrodynamic and magnetized runs.

The momentum constraint is,
\begin{equation}\label{eq:mom_con}
  \mathcal{M}^{i} = \tilde{D}_{j}(e^{6\phi}\tilde{A}^{ji})-\frac{2}{3}e^{6\phi}\tilde{D}^{i}K-8\pi e^{6\phi}S^{i}.
\end{equation}
In this case, we monitor the same $L_{2}$ norm as above for each of the three components of $\mathcal{M}^{i}$. We then take the norm of this vector: $||\mathcal{M}|| \equiv \sqrt{||\mathcal{M}^{x}||^{2} + ||\mathcal{M}^{y}||^{2} + ||\mathcal{M}^{z}||^{2}}$. This is plotted in Figure~\ref{fig:mom_con_comp}.

\begin{figure}[!tb]
  \includegraphics{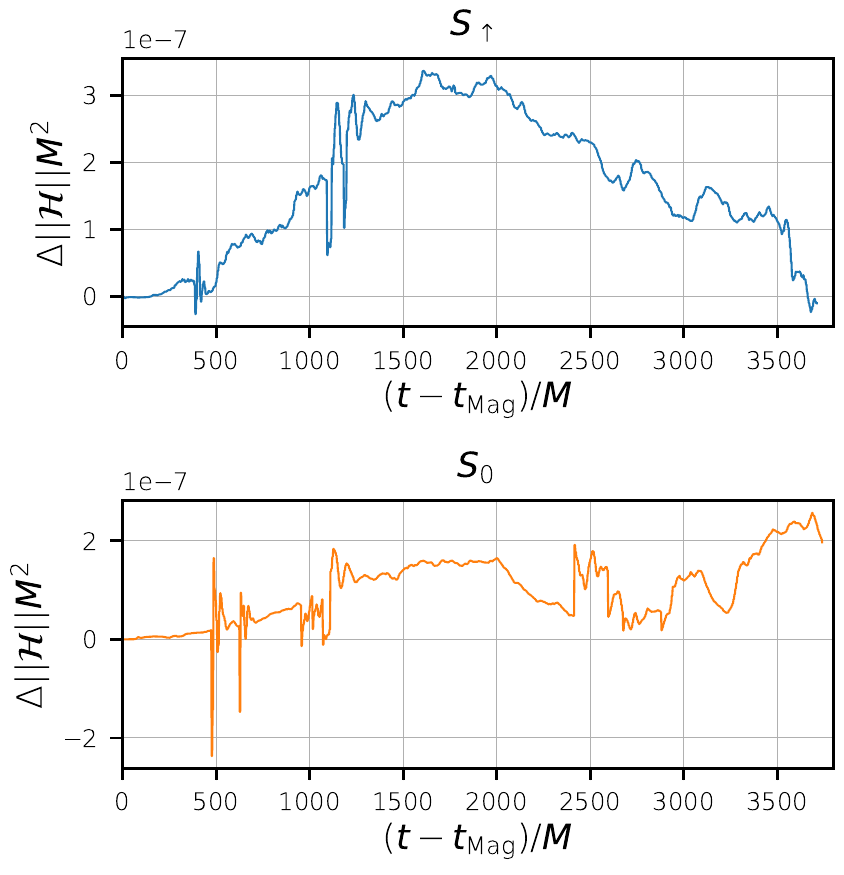}
  \caption{Evolution of the differences between the $L_{2}$ norms of the Hamiltonian constraint violations between the magnetized and non-magnetized evolutions, for both configurations.}\label{fig:ham_con_diff}
\end{figure}

\begin{figure}[!tb]
  \includegraphics{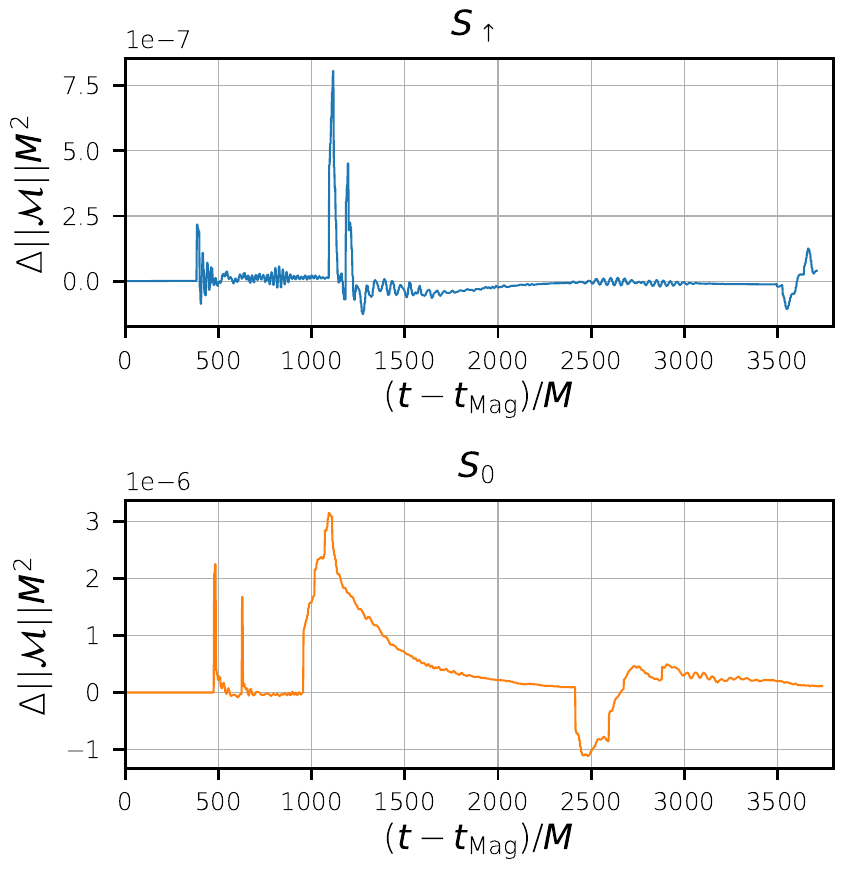}
  \caption{Evolution of the differences between the $L_{2}$ norms of the momentum constraint violations between the magnetized and non-magnetized evolutions, for both configurations.}\label{fig:mom_con_diff}
\end{figure}

As can be seen, the constraint values are nearly identical after field insertion, indicating that initial field has a small enough energy density to avoid introducing any noticable constraint violations. It is also clear that the differing dynamics between the two simulations have little impact on the constraint evolution. To verify that the near-indistinguishability of the two constraint evolutions between the magnetized and un-magnetized cases is not a mistake in our plot, Figure~\ref{fig:ham_con_diff} and Figure~\ref{fig:mom_con_diff} show the differences between the two sets of constraint violations for each configuration.

\FloatBarrier
\bibliography{Paper}

\begin{thebibliography}{91}
\expandafter\ifx\csname natexlab\endcsname\relax\def\natexlab#1{#1}\fi
\expandafter\ifx\csname bibnamefont\endcsname\relax
  \def\bibnamefont#1{#1}\fi
\expandafter\ifx\csname bibfnamefont\endcsname\relax
  \def\bibfnamefont#1{#1}\fi
\expandafter\ifx\csname citenamefont\endcsname\relax
  \def\citenamefont#1{#1}\fi
\expandafter\ifx\csname url\endcsname\relax
  \def\url#1{\texttt{#1}}\fi
\expandafter\ifx\csname urlprefix\endcsname\relax\def\urlprefix{URL }\fi
\providecommand{\bibinfo}[2]{#2}
\providecommand{\eprint}[2][]{\url{#2}}

\bibitem[{\citenamefont{Rees}(1984)}]{rees_black_1984}
\bibinfo{author}{\bibfnamefont{M.~J.} \bibnamefont{Rees}},
  \bibinfo{journal}{Annual Review of Astronomy and Astrophysics}
  \textbf{\bibinfo{volume}{22}}, \bibinfo{pages}{471} (\bibinfo{year}{1984}),
  ISSN \bibinfo{issn}{0066-4146}.

\bibitem[{\citenamefont{Shibata and Shapiro}(2002)}]{shibata_collapse_2002}
\bibinfo{author}{\bibfnamefont{M.}~\bibnamefont{Shibata}} \bibnamefont{and}
  \bibinfo{author}{\bibfnamefont{S.~L.} \bibnamefont{Shapiro}},
  \bibinfo{journal}{The Astrophysical Journal} \textbf{\bibinfo{volume}{572}},
  \bibinfo{pages}{L39} (\bibinfo{year}{2002}), ISSN \bibinfo{issn}{1538-4357}.

\bibitem[{\citenamefont{Lovelace et~al.}(2013)\citenamefont{Lovelace, Duez,
  Foucart, Kidder, Pfeiffer, Scheel, and Szil{\'a}gyi}}]{lovelace_massive_2013}
\bibinfo{author}{\bibfnamefont{G.}~\bibnamefont{Lovelace}},
  \bibinfo{author}{\bibfnamefont{M.~D.} \bibnamefont{Duez}},
  \bibinfo{author}{\bibfnamefont{F.}~\bibnamefont{Foucart}},
  \bibinfo{author}{\bibfnamefont{L.~E.} \bibnamefont{Kidder}},
  \bibinfo{author}{\bibfnamefont{H.~P.} \bibnamefont{Pfeiffer}},
  \bibinfo{author}{\bibfnamefont{M.~A.} \bibnamefont{Scheel}},
  \bibnamefont{and}
  \bibinfo{author}{\bibfnamefont{B.}~\bibnamefont{Szil{\'a}gyi}},
  \bibinfo{journal}{Classical and Quantum Gravity}
  \textbf{\bibinfo{volume}{30}}, \bibinfo{pages}{135004}
  (\bibinfo{year}{2013}), ISSN \bibinfo{issn}{0264-9381}.

\bibitem[{\citenamefont{Paschalidis
  et~al.}(2015{\natexlab{a}})\citenamefont{Paschalidis, Ruiz, and
  Shapiro}}]{paschalidis_relativistic_2015}
\bibinfo{author}{\bibfnamefont{V.}~\bibnamefont{Paschalidis}},
  \bibinfo{author}{\bibfnamefont{M.}~\bibnamefont{Ruiz}}, \bibnamefont{and}
  \bibinfo{author}{\bibfnamefont{S.~L.} \bibnamefont{Shapiro}},
  \bibinfo{journal}{The Astrophysical Journal Letters}
  \textbf{\bibinfo{volume}{806}}, \bibinfo{pages}{L14}
  (\bibinfo{year}{2015}{\natexlab{a}}),
  \urlprefix\url{https://dx.doi.org/10.1088/2041-8205/806/1/L14}.

\bibitem[{\citenamefont{Collaboration et~al.}(2020)\citenamefont{Collaboration,
  {the Virgo Collaboration}, {the KAGRA Collaboration}, Abbott, Abbott, Abbott,
  Abraham, Acernese, Ackley, Adams et~al.}}]{prospects_2020}
\bibinfo{author}{\bibfnamefont{T.~L.~S.} \bibnamefont{Collaboration}},
  \bibinfo{author}{\bibnamefont{{the Virgo Collaboration}}},
  \bibinfo{author}{\bibnamefont{{the KAGRA Collaboration}}},
  \bibinfo{author}{\bibfnamefont{B.~P.} \bibnamefont{Abbott}},
  \bibinfo{author}{\bibfnamefont{R.}~\bibnamefont{Abbott}},
  \bibinfo{author}{\bibfnamefont{T.~D.} \bibnamefont{Abbott}},
  \bibinfo{author}{\bibfnamefont{S.}~\bibnamefont{Abraham}},
  \bibinfo{author}{\bibfnamefont{F.}~\bibnamefont{Acernese}},
  \bibinfo{author}{\bibfnamefont{K.}~\bibnamefont{Ackley}},
  \bibinfo{author}{\bibfnamefont{C.}~\bibnamefont{Adams}},
  \bibnamefont{et~al.}, \bibinfo{journal}{Living Reviews in Relativity}
  \textbf{\bibinfo{volume}{23}}, \bibinfo{pages}{3} (\bibinfo{year}{2020}),
  ISSN \bibinfo{issn}{2367-3613, 1433-8351}, \eprint{1304.0670}.

\bibitem[{\citenamefont{Reitze et~al.}(2019)\citenamefont{Reitze, Adhikari,
  Ballmer, Barish, Barsotti, Billingsley, Brown, Chen, Coyne, Eisenstein
  et~al.}}]{reitze_cosmic_2019}
\bibinfo{author}{\bibfnamefont{D.}~\bibnamefont{Reitze}},
  \bibinfo{author}{\bibfnamefont{R.~X.} \bibnamefont{Adhikari}},
  \bibinfo{author}{\bibfnamefont{S.}~\bibnamefont{Ballmer}},
  \bibinfo{author}{\bibfnamefont{B.}~\bibnamefont{Barish}},
  \bibinfo{author}{\bibfnamefont{L.}~\bibnamefont{Barsotti}},
  \bibinfo{author}{\bibfnamefont{G.}~\bibnamefont{Billingsley}},
  \bibinfo{author}{\bibfnamefont{D.~A.} \bibnamefont{Brown}},
  \bibinfo{author}{\bibfnamefont{Y.}~\bibnamefont{Chen}},
  \bibinfo{author}{\bibfnamefont{D.}~\bibnamefont{Coyne}},
  \bibinfo{author}{\bibfnamefont{R.}~\bibnamefont{Eisenstein}},
  \bibnamefont{et~al.}, \emph{\bibinfo{title}{Cosmic {{Explorer}}: {{The
  U}}.{{S}}. {{Contribution}} to {{Gravitational-Wave Astronomy}} beyond
  {{LIGO}}}} (\bibinfo{year}{2019}), \eprint{arXiv:1907.04833}.

\bibitem[{\citenamefont{{Amaro-Seoane}
  et~al.}(2017)\citenamefont{{Amaro-Seoane}, Audley, Babak, Baker, Barausse,
  Bender, Berti, Binetruy, Born, Bortoluzzi et~al.}}]{amaro-seoane_laser_2017}
\bibinfo{author}{\bibfnamefont{P.}~\bibnamefont{{Amaro-Seoane}}},
  \bibinfo{author}{\bibfnamefont{H.}~\bibnamefont{Audley}},
  \bibinfo{author}{\bibfnamefont{S.}~\bibnamefont{Babak}},
  \bibinfo{author}{\bibfnamefont{J.}~\bibnamefont{Baker}},
  \bibinfo{author}{\bibfnamefont{E.}~\bibnamefont{Barausse}},
  \bibinfo{author}{\bibfnamefont{P.}~\bibnamefont{Bender}},
  \bibinfo{author}{\bibfnamefont{E.}~\bibnamefont{Berti}},
  \bibinfo{author}{\bibfnamefont{P.}~\bibnamefont{Binetruy}},
  \bibinfo{author}{\bibfnamefont{M.}~\bibnamefont{Born}},
  \bibinfo{author}{\bibfnamefont{D.}~\bibnamefont{Bortoluzzi}},
  \bibnamefont{et~al.}, \emph{\bibinfo{title}{Laser {{Interferometer Space
  Antenna}}}} (\bibinfo{year}{2017}), \eprint{arXiv:1702.00786}.

\bibitem[{\citenamefont{Sato et~al.}(2017)\citenamefont{Sato, Kawamura, Ando,
  Nakamura, Tsubono, Araya, Funaki, Ioka, Kanda, Moriwaki
  et~al.}}]{sato_status_2017}
\bibinfo{author}{\bibfnamefont{S.}~\bibnamefont{Sato}},
  \bibinfo{author}{\bibfnamefont{S.}~\bibnamefont{Kawamura}},
  \bibinfo{author}{\bibfnamefont{M.}~\bibnamefont{Ando}},
  \bibinfo{author}{\bibfnamefont{T.}~\bibnamefont{Nakamura}},
  \bibinfo{author}{\bibfnamefont{K.}~\bibnamefont{Tsubono}},
  \bibinfo{author}{\bibfnamefont{A.}~\bibnamefont{Araya}},
  \bibinfo{author}{\bibfnamefont{I.}~\bibnamefont{Funaki}},
  \bibinfo{author}{\bibfnamefont{K.}~\bibnamefont{Ioka}},
  \bibinfo{author}{\bibfnamefont{N.}~\bibnamefont{Kanda}},
  \bibinfo{author}{\bibfnamefont{S.}~\bibnamefont{Moriwaki}},
  \bibnamefont{et~al.}, \bibinfo{journal}{Journal of Physics: Conference
  Series} \textbf{\bibinfo{volume}{840}}, \bibinfo{pages}{012010}
  (\bibinfo{year}{2017}), ISSN \bibinfo{issn}{1742-6596}.

\bibitem[{\citenamefont{Papaloizou and
  Pringle}(1984)}]{papaloizou_dynamical_1984}
\bibinfo{author}{\bibfnamefont{J.~C.~B.} \bibnamefont{Papaloizou}}
  \bibnamefont{and} \bibinfo{author}{\bibfnamefont{J.~E.}
  \bibnamefont{Pringle}}, \bibinfo{journal}{Monthly Notices of the Royal
  Astronomical Society} \textbf{\bibinfo{volume}{208}}, \bibinfo{pages}{721}
  (\bibinfo{year}{1984}), ISSN \bibinfo{issn}{0035-8711}.

\bibitem[{\citenamefont{Hawley}(1987)}]{hawley_non-linear_1987}
\bibinfo{author}{\bibfnamefont{J.~F.} \bibnamefont{Hawley}},
  \bibinfo{journal}{Monthly Notices of the Royal Astronomical Society}
  \textbf{\bibinfo{volume}{225}}, \bibinfo{pages}{677} (\bibinfo{year}{1987}),
  ISSN \bibinfo{issn}{0035-8711}.

\bibitem[{\citenamefont{Blaes}(1985)}]{blaes_oscillations_1985}
\bibinfo{author}{\bibfnamefont{O.~M.} \bibnamefont{Blaes}},
  \bibinfo{journal}{Monthly Notices of the Royal Astronomical Society}
  \textbf{\bibinfo{volume}{216}}, \bibinfo{pages}{553} (\bibinfo{year}{1985}),
  ISSN \bibinfo{issn}{0035-8711}.

\bibitem[{\citenamefont{Blaes and Glatzel}(1986)}]{blaes_stability_1986}
\bibinfo{author}{\bibfnamefont{O.~M.} \bibnamefont{Blaes}} \bibnamefont{and}
  \bibinfo{author}{\bibfnamefont{W.}~\bibnamefont{Glatzel}},
  \bibinfo{journal}{Monthly Notices of the Royal Astronomical Society}
  \textbf{\bibinfo{volume}{220}}, \bibinfo{pages}{253} (\bibinfo{year}{1986}),
  ISSN \bibinfo{issn}{0035-8711}.

\bibitem[{\citenamefont{Goldreich et~al.}(1986)\citenamefont{Goldreich,
  Goodman, and Narayan}}]{goldreich_stability_1986}
\bibinfo{author}{\bibfnamefont{P.}~\bibnamefont{Goldreich}},
  \bibinfo{author}{\bibfnamefont{J.}~\bibnamefont{Goodman}}, \bibnamefont{and}
  \bibinfo{author}{\bibfnamefont{R.}~\bibnamefont{Narayan}},
  \bibinfo{journal}{Monthly Notices of the Royal Astronomical Society}
  \textbf{\bibinfo{volume}{221}}, \bibinfo{pages}{339} (\bibinfo{year}{1986}),
  ISSN \bibinfo{issn}{0035-8711}.

\bibitem[{\citenamefont{Narayan et~al.}(1987)\citenamefont{Narayan, Goldreich,
  and Goodman}}]{narayan_physics_1987}
\bibinfo{author}{\bibfnamefont{R.}~\bibnamefont{Narayan}},
  \bibinfo{author}{\bibfnamefont{P.}~\bibnamefont{Goldreich}},
  \bibnamefont{and} \bibinfo{author}{\bibfnamefont{J.}~\bibnamefont{Goodman}},
  \bibinfo{journal}{Monthly Notices of the Royal Astronomical Society}
  \textbf{\bibinfo{volume}{228}}, \bibinfo{pages}{1} (\bibinfo{year}{1987}),
  ISSN \bibinfo{issn}{0035-8711}.

\bibitem[{\citenamefont{Goodman and Narayan}(1988)}]{goodman_stability_1988}
\bibinfo{author}{\bibfnamefont{J.}~\bibnamefont{Goodman}} \bibnamefont{and}
  \bibinfo{author}{\bibfnamefont{R.}~\bibnamefont{Narayan}},
  \bibinfo{journal}{Monthly Notices of the Royal Astronomical Society}
  \textbf{\bibinfo{volume}{231}}, \bibinfo{pages}{97} (\bibinfo{year}{1988}),
  ISSN \bibinfo{issn}{0035-8711}.

\bibitem[{\citenamefont{Christodoulou and
  Narayan}(1992)}]{christodoulou_stability_1992}
\bibinfo{author}{\bibfnamefont{D.~M.} \bibnamefont{Christodoulou}}
  \bibnamefont{and} \bibinfo{author}{\bibfnamefont{R.}~\bibnamefont{Narayan}},
  \bibinfo{journal}{The Astrophysical Journal} \textbf{\bibinfo{volume}{388}},
  \bibinfo{pages}{451} (\bibinfo{year}{1992}), ISSN \bibinfo{issn}{0004-637X}.

\bibitem[{\citenamefont{Papaloizou and
  Pringle}(1985)}]{papaloizou_dynamical_1985}
\bibinfo{author}{\bibfnamefont{J.~C.~B.} \bibnamefont{Papaloizou}}
  \bibnamefont{and} \bibinfo{author}{\bibfnamefont{J.~E.}
  \bibnamefont{Pringle}}, \bibinfo{journal}{Monthly Notices of the Royal
  Astronomical Society} \textbf{\bibinfo{volume}{213}}, \bibinfo{pages}{799}
  (\bibinfo{year}{1985}), ISSN \bibinfo{issn}{0035-8711}.

\bibitem[{\citenamefont{Strutt}(1917)}]{strutt_dynamics_1917}
\bibinfo{author}{\bibfnamefont{J.~W.} \bibnamefont{Strutt}},
  \bibinfo{journal}{Proceedings of the Royal Society of London. Series A,
  Containing Papers of a Mathematical and Physical Character}
  \textbf{\bibinfo{volume}{93}}, \bibinfo{pages}{148} (\bibinfo{year}{1917}).

\bibitem[{\citenamefont{Kojima}(1986)}]{kojima_dynamical_1986}
\bibinfo{author}{\bibfnamefont{Y.}~\bibnamefont{Kojima}},
  \bibinfo{journal}{Progress of Theoretical Physics}
  \textbf{\bibinfo{volume}{75}}, \bibinfo{pages}{1464} (\bibinfo{year}{1986}),
  ISSN \bibinfo{issn}{0033-068X}.

\bibitem[{\citenamefont{Korobkin et~al.}(2011)\citenamefont{Korobkin,
  Abdikamalov, Schnetter, Stergioulas, and Zink}}]{korobkin_stability_2011}
\bibinfo{author}{\bibfnamefont{O.}~\bibnamefont{Korobkin}},
  \bibinfo{author}{\bibfnamefont{E.~B.} \bibnamefont{Abdikamalov}},
  \bibinfo{author}{\bibfnamefont{E.}~\bibnamefont{Schnetter}},
  \bibinfo{author}{\bibfnamefont{N.}~\bibnamefont{Stergioulas}},
  \bibnamefont{and} \bibinfo{author}{\bibfnamefont{B.}~\bibnamefont{Zink}},
  \bibinfo{journal}{Physical Review D} \textbf{\bibinfo{volume}{83}},
  \bibinfo{pages}{043007} (\bibinfo{year}{2011}).

\bibitem[{\citenamefont{Kiuchi et~al.}(2011)\citenamefont{Kiuchi, Shibata,
  Montero, and Font}}]{kiuchi_gravitational_2011}
\bibinfo{author}{\bibfnamefont{K.}~\bibnamefont{Kiuchi}},
  \bibinfo{author}{\bibfnamefont{M.}~\bibnamefont{Shibata}},
  \bibinfo{author}{\bibfnamefont{P.~J.} \bibnamefont{Montero}},
  \bibnamefont{and} \bibinfo{author}{\bibfnamefont{J.~A.} \bibnamefont{Font}},
  \bibinfo{journal}{Physical Review Letters} \textbf{\bibinfo{volume}{106}},
  \bibinfo{pages}{251102} (\bibinfo{year}{2011}).

\bibitem[{\citenamefont{Nealon et~al.}(2018)\citenamefont{Nealon, Price,
  Bonnerot, and Lodato}}]{nealon_papaloizoupringle_2018}
\bibinfo{author}{\bibfnamefont{R.}~\bibnamefont{Nealon}},
  \bibinfo{author}{\bibfnamefont{D.~J.} \bibnamefont{Price}},
  \bibinfo{author}{\bibfnamefont{C.}~\bibnamefont{Bonnerot}}, \bibnamefont{and}
  \bibinfo{author}{\bibfnamefont{G.}~\bibnamefont{Lodato}},
  \bibinfo{journal}{Monthly Notices of the Royal Astronomical Society}
  \textbf{\bibinfo{volume}{474}}, \bibinfo{pages}{1737} (\bibinfo{year}{2018}),
  ISSN \bibinfo{issn}{0035-8711}.

\bibitem[{\citenamefont{Toscani et~al.}(2019)\citenamefont{Toscani, Lodato, and
  Nealon}}]{toscani_gravitational_2019}
\bibinfo{author}{\bibfnamefont{M.}~\bibnamefont{Toscani}},
  \bibinfo{author}{\bibfnamefont{G.}~\bibnamefont{Lodato}}, \bibnamefont{and}
  \bibinfo{author}{\bibfnamefont{R.}~\bibnamefont{Nealon}},
  \bibinfo{journal}{Monthly Notices of the Royal Astronomical Society}
  \textbf{\bibinfo{volume}{489}}, \bibinfo{pages}{699} (\bibinfo{year}{2019}),
  ISSN \bibinfo{issn}{0035-8711}.

\bibitem[{\citenamefont{Mewes et~al.}(2016)\citenamefont{Mewes, Font, Galeazzi,
  Montero, and Stergioulas}}]{mewes_numerical_2016}
\bibinfo{author}{\bibfnamefont{V.}~\bibnamefont{Mewes}},
  \bibinfo{author}{\bibfnamefont{J.~A.} \bibnamefont{Font}},
  \bibinfo{author}{\bibfnamefont{F.}~\bibnamefont{Galeazzi}},
  \bibinfo{author}{\bibfnamefont{P.~J.} \bibnamefont{Montero}},
  \bibnamefont{and}
  \bibinfo{author}{\bibfnamefont{N.}~\bibnamefont{Stergioulas}},
  \bibinfo{journal}{Physical Review D} \textbf{\bibinfo{volume}{93}},
  \bibinfo{pages}{064055} (\bibinfo{year}{2016}).

\bibitem[{\citenamefont{Wessel et~al.}(2021)\citenamefont{Wessel, Paschalidis,
  Tsokaros, Ruiz, and Shapiro}}]{wessel_gravitational_2021}
\bibinfo{author}{\bibfnamefont{E.}~\bibnamefont{Wessel}},
  \bibinfo{author}{\bibfnamefont{V.}~\bibnamefont{Paschalidis}},
  \bibinfo{author}{\bibfnamefont{A.}~\bibnamefont{Tsokaros}},
  \bibinfo{author}{\bibfnamefont{M.}~\bibnamefont{Ruiz}}, \bibnamefont{and}
  \bibinfo{author}{\bibfnamefont{S.~L.} \bibnamefont{Shapiro}},
  \bibinfo{journal}{Physical Review D} \textbf{\bibinfo{volume}{103}},
  \bibinfo{pages}{043013} (\bibinfo{year}{2021}).

\bibitem[{\citenamefont{Tsokaros et~al.}(2022)\citenamefont{Tsokaros, Ruiz,
  Shapiro, and Paschalidis}}]{tsokaros_self-gravitating_2022}
\bibinfo{author}{\bibfnamefont{A.}~\bibnamefont{Tsokaros}},
  \bibinfo{author}{\bibfnamefont{M.}~\bibnamefont{Ruiz}},
  \bibinfo{author}{\bibfnamefont{S.~L.} \bibnamefont{Shapiro}},
  \bibnamefont{and}
  \bibinfo{author}{\bibfnamefont{V.}~\bibnamefont{Paschalidis}},
  \bibinfo{journal}{Physical Review D} \textbf{\bibinfo{volume}{106}},
  \bibinfo{pages}{104010} (\bibinfo{year}{2022}).

\bibitem[{\citenamefont{Shibata et~al.}(2021)\citenamefont{Shibata, Kiuchi,
  Fujibayashi, and Sekiguchi}}]{shibata_alternative_2021}
\bibinfo{author}{\bibfnamefont{M.}~\bibnamefont{Shibata}},
  \bibinfo{author}{\bibfnamefont{K.}~\bibnamefont{Kiuchi}},
  \bibinfo{author}{\bibfnamefont{S.}~\bibnamefont{Fujibayashi}},
  \bibnamefont{and}
  \bibinfo{author}{\bibfnamefont{Y.}~\bibnamefont{Sekiguchi}},
  \bibinfo{journal}{Physical Review D} \textbf{\bibinfo{volume}{103}},
  \bibinfo{pages}{063037} (\bibinfo{year}{2021}).

\bibitem[{\citenamefont{Balbus and Hawley}(1992)}]{balbus_powerful_1992}
\bibinfo{author}{\bibfnamefont{S.~A.} \bibnamefont{Balbus}} \bibnamefont{and}
  \bibinfo{author}{\bibfnamefont{J.~F.} \bibnamefont{Hawley}},
  \bibinfo{journal}{The Astrophysical Journal} \textbf{\bibinfo{volume}{400}},
  \bibinfo{pages}{610} (\bibinfo{year}{1992}), ISSN \bibinfo{issn}{0004-637X}.

\bibitem[{\citenamefont{Balbus and Hawley}(1998)}]{balbus_instability_1998}
\bibinfo{author}{\bibfnamefont{S.~A.} \bibnamefont{Balbus}} \bibnamefont{and}
  \bibinfo{author}{\bibfnamefont{J.~F.} \bibnamefont{Hawley}},
  \bibinfo{journal}{Reviews of Modern Physics} \textbf{\bibinfo{volume}{70}},
  \bibinfo{pages}{1} (\bibinfo{year}{1998}).

\bibitem[{\citenamefont{Bugli et~al.}(2018)\citenamefont{Bugli, Guilet,
  M{\"u}ller, Del~Zanna, Bucciantini, and
  Montero}}]{bugli_papaloizoupringle_2018}
\bibinfo{author}{\bibfnamefont{M.}~\bibnamefont{Bugli}},
  \bibinfo{author}{\bibfnamefont{J.}~\bibnamefont{Guilet}},
  \bibinfo{author}{\bibfnamefont{E.}~\bibnamefont{M{\"u}ller}},
  \bibinfo{author}{\bibfnamefont{L.}~\bibnamefont{Del~Zanna}},
  \bibinfo{author}{\bibfnamefont{N.}~\bibnamefont{Bucciantini}},
  \bibnamefont{and} \bibinfo{author}{\bibfnamefont{P.~J.}
  \bibnamefont{Montero}}, \bibinfo{journal}{Monthly Notices of the Royal
  Astronomical Society} \textbf{\bibinfo{volume}{475}}, \bibinfo{pages}{108}
  (\bibinfo{year}{2018}), ISSN \bibinfo{issn}{0035-8711}.

\bibitem[{\citenamefont{Christodoulou}(1970)}]{christodoulou_reversible_1970}
\bibinfo{author}{\bibfnamefont{D.}~\bibnamefont{Christodoulou}},
  \bibinfo{journal}{Physical Review Letters} \textbf{\bibinfo{volume}{25}},
  \bibinfo{pages}{1596} (\bibinfo{year}{1970}).

\bibitem[{\citenamefont{Bozzola}(2021)}]{bozzola_kuibit_2021}
\bibinfo{author}{\bibfnamefont{G.}~\bibnamefont{Bozzola}},
  \bibinfo{journal}{Journal of Open Source Software}
  \textbf{\bibinfo{volume}{6}}, \bibinfo{pages}{3099} (\bibinfo{year}{2021}),
  ISSN \bibinfo{issn}{2475-9066}.

\bibitem[{\citenamefont{Tsokaros et~al.}(2019)\citenamefont{Tsokaros, Ury{\=u},
  and Shapiro}}]{tsokaros_complete_2019}
\bibinfo{author}{\bibfnamefont{A.}~\bibnamefont{Tsokaros}},
  \bibinfo{author}{\bibfnamefont{K.}~\bibnamefont{Ury{\=u}}}, \bibnamefont{and}
  \bibinfo{author}{\bibfnamefont{S.~L.} \bibnamefont{Shapiro}},
  \bibinfo{journal}{Physical Review D} \textbf{\bibinfo{volume}{99}},
  \bibinfo{pages}{041501} (\bibinfo{year}{2019}).

\bibitem[{\citenamefont{Komatsu et~al.}(1989)\citenamefont{Komatsu, Eriguchi,
  and Hachisu}}]{komatsu_rapidly_1989}
\bibinfo{author}{\bibfnamefont{H.}~\bibnamefont{Komatsu}},
  \bibinfo{author}{\bibfnamefont{Y.}~\bibnamefont{Eriguchi}}, \bibnamefont{and}
  \bibinfo{author}{\bibfnamefont{I.}~\bibnamefont{Hachisu}},
  \bibinfo{journal}{Monthly Notices of the Royal Astronomical Society}
  \textbf{\bibinfo{volume}{237}}, \bibinfo{pages}{355} (\bibinfo{year}{1989}),
  ISSN \bibinfo{issn}{0035-8711}.

\bibitem[{\citenamefont{Tsokaros and Ury{\=u}}(2007)}]{tsokaros_numerical_2007}
\bibinfo{author}{\bibfnamefont{A.~A.} \bibnamefont{Tsokaros}} \bibnamefont{and}
  \bibinfo{author}{\bibfnamefont{K.}~\bibnamefont{Ury{\=u}}},
  \bibinfo{journal}{Physical Review D} \textbf{\bibinfo{volume}{75}},
  \bibinfo{pages}{044026} (\bibinfo{year}{2007}).

\bibitem[{\citenamefont{Toomre}(1964)}]{toomre_gravitational_1964-1}
\bibinfo{author}{\bibfnamefont{A.}~\bibnamefont{Toomre}}, \bibinfo{journal}{The
  Astrophysical Journal} \textbf{\bibinfo{volume}{139}}, \bibinfo{pages}{1217}
  (\bibinfo{year}{1964}), ISSN \bibinfo{issn}{0004-637X}.

\bibitem[{\citenamefont{Etienne et~al.}(2007)\citenamefont{Etienne, Faber, Liu,
  Shapiro, and Baumgarte}}]{etienne_filling_2007}
\bibinfo{author}{\bibfnamefont{Z.~B.} \bibnamefont{Etienne}},
  \bibinfo{author}{\bibfnamefont{J.~A.} \bibnamefont{Faber}},
  \bibinfo{author}{\bibfnamefont{Y.~T.} \bibnamefont{Liu}},
  \bibinfo{author}{\bibfnamefont{S.~L.} \bibnamefont{Shapiro}},
  \bibnamefont{and} \bibinfo{author}{\bibfnamefont{T.~W.}
  \bibnamefont{Baumgarte}}, \bibinfo{journal}{Physical Review D}
  \textbf{\bibinfo{volume}{76}}, \bibinfo{pages}{101503}
  (\bibinfo{year}{2007}).

\bibitem[{\citenamefont{Duez et~al.}(2005)\citenamefont{Duez, Liu, Shapiro, and
  Stephens}}]{duez_relativistic_2005}
\bibinfo{author}{\bibfnamefont{M.~D.} \bibnamefont{Duez}},
  \bibinfo{author}{\bibfnamefont{Y.~T.} \bibnamefont{Liu}},
  \bibinfo{author}{\bibfnamefont{S.~L.} \bibnamefont{Shapiro}},
  \bibnamefont{and} \bibinfo{author}{\bibfnamefont{B.~C.}
  \bibnamefont{Stephens}}, \bibinfo{journal}{Physical Review D}
  \textbf{\bibinfo{volume}{72}}, \bibinfo{pages}{024028}
  (\bibinfo{year}{2005}).

\bibitem[{\citenamefont{Etienne et~al.}(2010)\citenamefont{Etienne, Liu, and
  Shapiro}}]{etienne_relativistic_2010}
\bibinfo{author}{\bibfnamefont{Z.~B.} \bibnamefont{Etienne}},
  \bibinfo{author}{\bibfnamefont{Y.~T.} \bibnamefont{Liu}}, \bibnamefont{and}
  \bibinfo{author}{\bibfnamefont{S.~L.} \bibnamefont{Shapiro}},
  \bibinfo{journal}{Physical Review D} \textbf{\bibinfo{volume}{82}},
  \bibinfo{pages}{084031} (\bibinfo{year}{2010}).

\bibitem[{\citenamefont{Etienne et~al.}(2012)\citenamefont{Etienne, Liu,
  Paschalidis, and Shapiro}}]{etienne_general_2012}
\bibinfo{author}{\bibfnamefont{Z.~B.} \bibnamefont{Etienne}},
  \bibinfo{author}{\bibfnamefont{Y.~T.} \bibnamefont{Liu}},
  \bibinfo{author}{\bibfnamefont{V.}~\bibnamefont{Paschalidis}},
  \bibnamefont{and} \bibinfo{author}{\bibfnamefont{S.~L.}
  \bibnamefont{Shapiro}}, \bibinfo{journal}{Physical Review D}
  \textbf{\bibinfo{volume}{85}}, \bibinfo{pages}{064029}
  (\bibinfo{year}{2012}).

\bibitem[{noa()}]{noauthor_cactus_nodate-1}
\emph{\bibinfo{title}{Cactus}},
  \bibinfo{howpublished}{https://www.cactuscode.org/}.

\bibitem[{\citenamefont{Schnetter et~al.}(2004)\citenamefont{Schnetter, Hawley,
  and Hawke}}]{schnetter_evolutions_2004}
\bibinfo{author}{\bibfnamefont{E.}~\bibnamefont{Schnetter}},
  \bibinfo{author}{\bibfnamefont{S.~H.} \bibnamefont{Hawley}},
  \bibnamefont{and} \bibinfo{author}{\bibfnamefont{I.}~\bibnamefont{Hawke}},
  \bibinfo{journal}{Classical and Quantum Gravity}
  \textbf{\bibinfo{volume}{21}}, \bibinfo{pages}{1465} (\bibinfo{year}{2004}),
  ISSN \bibinfo{issn}{0264-9381, 1361-6382}, \eprint{gr-qc/0310042}.

\bibitem[{\citenamefont{Etienne et~al.}(2015)\citenamefont{Etienne,
  Paschalidis, Haas, M{\"o}sta, and Shapiro}}]{etienne_illinoisgrmhd_2015}
\bibinfo{author}{\bibfnamefont{Z.~B.} \bibnamefont{Etienne}},
  \bibinfo{author}{\bibfnamefont{V.}~\bibnamefont{Paschalidis}},
  \bibinfo{author}{\bibfnamefont{R.}~\bibnamefont{Haas}},
  \bibinfo{author}{\bibfnamefont{P.}~\bibnamefont{M{\"o}sta}},
  \bibnamefont{and} \bibinfo{author}{\bibfnamefont{S.~L.}
  \bibnamefont{Shapiro}}, \bibinfo{journal}{Classical and Quantum Gravity}
  \textbf{\bibinfo{volume}{32}}, \bibinfo{pages}{175009}
  (\bibinfo{year}{2015}), ISSN \bibinfo{issn}{0264-9381}.

\bibitem[{\citenamefont{Shibata and Nakamura}(1995)}]{shibata_evolution_1995}
\bibinfo{author}{\bibfnamefont{M.}~\bibnamefont{Shibata}} \bibnamefont{and}
  \bibinfo{author}{\bibfnamefont{T.}~\bibnamefont{Nakamura}},
  \bibinfo{journal}{Physical Review D} \textbf{\bibinfo{volume}{52}},
  \bibinfo{pages}{5428} (\bibinfo{year}{1995}).

\bibitem[{\citenamefont{Baumgarte and
  Shapiro}(1998)}]{baumgarte_numerical_1998}
\bibinfo{author}{\bibfnamefont{T.~W.} \bibnamefont{Baumgarte}}
  \bibnamefont{and} \bibinfo{author}{\bibfnamefont{S.~L.}
  \bibnamefont{Shapiro}}, \bibinfo{journal}{Physical Review D}
  \textbf{\bibinfo{volume}{59}}, \bibinfo{pages}{024007}
  (\bibinfo{year}{1998}).

\bibitem[{\citenamefont{Campanelli et~al.}(2006)\citenamefont{Campanelli,
  Lousto, Marronetti, and Zlochower}}]{campanelli_accurate_2006}
\bibinfo{author}{\bibfnamefont{M.}~\bibnamefont{Campanelli}},
  \bibinfo{author}{\bibfnamefont{C.~O.} \bibnamefont{Lousto}},
  \bibinfo{author}{\bibfnamefont{P.}~\bibnamefont{Marronetti}},
  \bibnamefont{and}
  \bibinfo{author}{\bibfnamefont{Y.}~\bibnamefont{Zlochower}},
  \bibinfo{journal}{Physical Review Letters} \textbf{\bibinfo{volume}{96}},
  \bibinfo{pages}{111101} (\bibinfo{year}{2006}).

\bibitem[{\citenamefont{Baker et~al.}(2006)\citenamefont{Baker, Centrella,
  Choi, Koppitz, and {van Meter}}}]{baker_gravitational-wave_2006}
\bibinfo{author}{\bibfnamefont{J.~G.} \bibnamefont{Baker}},
  \bibinfo{author}{\bibfnamefont{J.}~\bibnamefont{Centrella}},
  \bibinfo{author}{\bibfnamefont{D.-I.} \bibnamefont{Choi}},
  \bibinfo{author}{\bibfnamefont{M.}~\bibnamefont{Koppitz}}, \bibnamefont{and}
  \bibinfo{author}{\bibfnamefont{J.}~\bibnamefont{{van Meter}}},
  \bibinfo{journal}{Physical Review Letters} \textbf{\bibinfo{volume}{96}},
  \bibinfo{pages}{111102} (\bibinfo{year}{2006}).

\bibitem[{\citenamefont{Bona et~al.}(1995)\citenamefont{Bona, Mass{\'o},
  Seidel, and Stela}}]{bona_new_1995}
\bibinfo{author}{\bibfnamefont{C.}~\bibnamefont{Bona}},
  \bibinfo{author}{\bibfnamefont{J.}~\bibnamefont{Mass{\'o}}},
  \bibinfo{author}{\bibfnamefont{E.}~\bibnamefont{Seidel}}, \bibnamefont{and}
  \bibinfo{author}{\bibfnamefont{J.}~\bibnamefont{Stela}},
  \bibinfo{journal}{Physical Review Letters} \textbf{\bibinfo{volume}{75}},
  \bibinfo{pages}{600} (\bibinfo{year}{1995}).

\bibitem[{\citenamefont{Alcubierre et~al.}(2003)\citenamefont{Alcubierre,
  Br{\"u}gmann, Diener, Koppitz, Pollney, Seidel, and
  Takahashi}}]{alcubierre_gauge_2003}
\bibinfo{author}{\bibfnamefont{M.}~\bibnamefont{Alcubierre}},
  \bibinfo{author}{\bibfnamefont{B.}~\bibnamefont{Br{\"u}gmann}},
  \bibinfo{author}{\bibfnamefont{P.}~\bibnamefont{Diener}},
  \bibinfo{author}{\bibfnamefont{M.}~\bibnamefont{Koppitz}},
  \bibinfo{author}{\bibfnamefont{D.}~\bibnamefont{Pollney}},
  \bibinfo{author}{\bibfnamefont{E.}~\bibnamefont{Seidel}}, \bibnamefont{and}
  \bibinfo{author}{\bibfnamefont{R.}~\bibnamefont{Takahashi}},
  \bibinfo{journal}{Physical Review D} \textbf{\bibinfo{volume}{67}},
  \bibinfo{pages}{084023} (\bibinfo{year}{2003}).

\bibitem[{\citenamefont{Hinder et~al.}(2013)\citenamefont{Hinder, Buonanno,
  Boyle, Etienne, Healy, {Johnson-McDaniel}, Nagar, Nakano, Pan, Pfeiffer
  et~al.}}]{hinder_error-analysis_2013}
\bibinfo{author}{\bibfnamefont{I.}~\bibnamefont{Hinder}},
  \bibinfo{author}{\bibfnamefont{A.}~\bibnamefont{Buonanno}},
  \bibinfo{author}{\bibfnamefont{M.}~\bibnamefont{Boyle}},
  \bibinfo{author}{\bibfnamefont{Z.~B.} \bibnamefont{Etienne}},
  \bibinfo{author}{\bibfnamefont{J.}~\bibnamefont{Healy}},
  \bibinfo{author}{\bibfnamefont{N.~K.} \bibnamefont{{Johnson-McDaniel}}},
  \bibinfo{author}{\bibfnamefont{A.}~\bibnamefont{Nagar}},
  \bibinfo{author}{\bibfnamefont{H.}~\bibnamefont{Nakano}},
  \bibinfo{author}{\bibfnamefont{Y.}~\bibnamefont{Pan}},
  \bibinfo{author}{\bibfnamefont{H.~P.} \bibnamefont{Pfeiffer}},
  \bibnamefont{et~al.}, \bibinfo{journal}{Classical and Quantum Gravity}
  \textbf{\bibinfo{volume}{31}}, \bibinfo{pages}{025012}
  (\bibinfo{year}{2013}), ISSN \bibinfo{issn}{0264-9381}.

\bibitem[{\citenamefont{Hawley et~al.}(2011)\citenamefont{Hawley, Guan, and
  Krolik}}]{hawley_assessing_2011}
\bibinfo{author}{\bibfnamefont{J.~F.} \bibnamefont{Hawley}},
  \bibinfo{author}{\bibfnamefont{X.}~\bibnamefont{Guan}}, \bibnamefont{and}
  \bibinfo{author}{\bibfnamefont{J.~H.} \bibnamefont{Krolik}},
  \bibinfo{journal}{The Astrophysical Journal} \textbf{\bibinfo{volume}{738}},
  \bibinfo{pages}{84} (\bibinfo{year}{2011}), ISSN \bibinfo{issn}{0004-637X}.

\bibitem[{\citenamefont{Paschalidis}(2017)}]{paschalidis_general_2017}
\bibinfo{author}{\bibfnamefont{V.}~\bibnamefont{Paschalidis}},
  \bibinfo{journal}{Classical and Quantum Gravity}
  \textbf{\bibinfo{volume}{34}}, \bibinfo{pages}{084002}
  (\bibinfo{year}{2017}), ISSN \bibinfo{issn}{0264-9381}.

\bibitem[{\citenamefont{Gold et~al.}(2014)\citenamefont{Gold, Paschalidis,
  Etienne, Shapiro, and Pfeiffer}}]{gold_accretion_2014}
\bibinfo{author}{\bibfnamefont{R.}~\bibnamefont{Gold}},
  \bibinfo{author}{\bibfnamefont{V.}~\bibnamefont{Paschalidis}},
  \bibinfo{author}{\bibfnamefont{Z.~B.} \bibnamefont{Etienne}},
  \bibinfo{author}{\bibfnamefont{S.~L.} \bibnamefont{Shapiro}},
  \bibnamefont{and} \bibinfo{author}{\bibfnamefont{H.~P.}
  \bibnamefont{Pfeiffer}}, \bibinfo{journal}{Physical Review D}
  \textbf{\bibinfo{volume}{89}}, \bibinfo{pages}{064060}
  (\bibinfo{year}{2014}).

\bibitem[{\citenamefont{Paschalidis et~al.}(2021)\citenamefont{Paschalidis,
  Bright, Ruiz, and Gold}}]{paschalidis_minidisk_2021}
\bibinfo{author}{\bibfnamefont{V.}~\bibnamefont{Paschalidis}},
  \bibinfo{author}{\bibfnamefont{J.}~\bibnamefont{Bright}},
  \bibinfo{author}{\bibfnamefont{M.}~\bibnamefont{Ruiz}}, \bibnamefont{and}
  \bibinfo{author}{\bibfnamefont{R.}~\bibnamefont{Gold}}, \bibinfo{journal}{The
  Astrophysical Journal Letters} \textbf{\bibinfo{volume}{910}},
  \bibinfo{pages}{L26} (\bibinfo{year}{2021}), ISSN \bibinfo{issn}{2041-8205}.

\bibitem[{\citenamefont{Farris et~al.}(2012)\citenamefont{Farris, Gold,
  Paschalidis, Etienne, and Shapiro}}]{farris_binary_2012}
\bibinfo{author}{\bibfnamefont{B.~D.} \bibnamefont{Farris}},
  \bibinfo{author}{\bibfnamefont{R.}~\bibnamefont{Gold}},
  \bibinfo{author}{\bibfnamefont{V.}~\bibnamefont{Paschalidis}},
  \bibinfo{author}{\bibfnamefont{Z.~B.} \bibnamefont{Etienne}},
  \bibnamefont{and} \bibinfo{author}{\bibfnamefont{S.~L.}
  \bibnamefont{Shapiro}}, \bibinfo{journal}{Physical Review Letters}
  \textbf{\bibinfo{volume}{109}}, \bibinfo{pages}{221102}
  (\bibinfo{year}{2012}).

\bibitem[{\citenamefont{Etienne et~al.}(2008)\citenamefont{Etienne, Faber, Liu,
  Shapiro, Taniguchi, and Baumgarte}}]{etienne_fully_2008}
\bibinfo{author}{\bibfnamefont{Z.~B.} \bibnamefont{Etienne}},
  \bibinfo{author}{\bibfnamefont{J.~A.} \bibnamefont{Faber}},
  \bibinfo{author}{\bibfnamefont{Y.~T.} \bibnamefont{Liu}},
  \bibinfo{author}{\bibfnamefont{S.~L.} \bibnamefont{Shapiro}},
  \bibinfo{author}{\bibfnamefont{K.}~\bibnamefont{Taniguchi}},
  \bibnamefont{and} \bibinfo{author}{\bibfnamefont{T.~W.}
  \bibnamefont{Baumgarte}}, \bibinfo{journal}{Physical Review D}
  \textbf{\bibinfo{volume}{77}}, \bibinfo{pages}{084002}
  (\bibinfo{year}{2008}).

\bibitem[{\citenamefont{Paschalidis
  et~al.}(2015{\natexlab{b}})\citenamefont{Paschalidis, East, Pretorius, and
  Shapiro}}]{paschalidis_one-arm_2015}
\bibinfo{author}{\bibfnamefont{V.}~\bibnamefont{Paschalidis}},
  \bibinfo{author}{\bibfnamefont{W.~E.} \bibnamefont{East}},
  \bibinfo{author}{\bibfnamefont{F.}~\bibnamefont{Pretorius}},
  \bibnamefont{and} \bibinfo{author}{\bibfnamefont{S.~L.}
  \bibnamefont{Shapiro}}, \bibinfo{journal}{Physical Review D}
  \textbf{\bibinfo{volume}{92}}, \bibinfo{pages}{121502}
  (\bibinfo{year}{2015}{\natexlab{b}}).

\bibitem[{\citenamefont{East et~al.}(2016{\natexlab{a}})\citenamefont{East,
  Paschalidis, and Pretorius}}]{east_equation_2016}
\bibinfo{author}{\bibfnamefont{W.~E.} \bibnamefont{East}},
  \bibinfo{author}{\bibfnamefont{V.}~\bibnamefont{Paschalidis}},
  \bibnamefont{and}
  \bibinfo{author}{\bibfnamefont{F.}~\bibnamefont{Pretorius}},
  \bibinfo{journal}{Classical and Quantum Gravity}
  \textbf{\bibinfo{volume}{33}}, \bibinfo{pages}{244004}
  (\bibinfo{year}{2016}{\natexlab{a}}), ISSN \bibinfo{issn}{0264-9381}.

\bibitem[{\citenamefont{East et~al.}(2016{\natexlab{b}})\citenamefont{East,
  Paschalidis, Pretorius, and Shapiro}}]{east_relativistic_2016}
\bibinfo{author}{\bibfnamefont{W.~E.} \bibnamefont{East}},
  \bibinfo{author}{\bibfnamefont{V.}~\bibnamefont{Paschalidis}},
  \bibinfo{author}{\bibfnamefont{F.}~\bibnamefont{Pretorius}},
  \bibnamefont{and} \bibinfo{author}{\bibfnamefont{S.~L.}
  \bibnamefont{Shapiro}}, \bibinfo{journal}{Physical Review D}
  \textbf{\bibinfo{volume}{93}}, \bibinfo{pages}{024011}
  (\bibinfo{year}{2016}{\natexlab{b}}).

\bibitem[{\citenamefont{Hawley et~al.}(2013)\citenamefont{Hawley, Richers,
  Guan, and Krolik}}]{hawley_testing_2013}
\bibinfo{author}{\bibfnamefont{J.~F.} \bibnamefont{Hawley}},
  \bibinfo{author}{\bibfnamefont{S.~A.} \bibnamefont{Richers}},
  \bibinfo{author}{\bibfnamefont{X.}~\bibnamefont{Guan}}, \bibnamefont{and}
  \bibinfo{author}{\bibfnamefont{J.~H.} \bibnamefont{Krolik}},
  \bibinfo{journal}{The Astrophysical Journal} \textbf{\bibinfo{volume}{772}},
  \bibinfo{pages}{102} (\bibinfo{year}{2013}), ISSN \bibinfo{issn}{0004-637X,
  1538-4357}, \eprint{1306.0243}.

\bibitem[{\citenamefont{Reisswig and Pollney}(2011)}]{reisswig_notes_2011}
\bibinfo{author}{\bibfnamefont{C.}~\bibnamefont{Reisswig}} \bibnamefont{and}
  \bibinfo{author}{\bibfnamefont{D.}~\bibnamefont{Pollney}},
  \bibinfo{journal}{Classical and Quantum Gravity}
  \textbf{\bibinfo{volume}{28}}, \bibinfo{pages}{195015}
  (\bibinfo{year}{2011}), ISSN \bibinfo{issn}{0264-9381}.

\bibitem[{\citenamefont{Espino et~al.}(2019)\citenamefont{Espino, Paschalidis,
  Baumgarte, and Shapiro}}]{espino_dynamical_2019}
\bibinfo{author}{\bibfnamefont{P.~L.} \bibnamefont{Espino}},
  \bibinfo{author}{\bibfnamefont{V.}~\bibnamefont{Paschalidis}},
  \bibinfo{author}{\bibfnamefont{T.~W.} \bibnamefont{Baumgarte}},
  \bibnamefont{and} \bibinfo{author}{\bibfnamefont{S.~L.}
  \bibnamefont{Shapiro}}, \bibinfo{journal}{Physical Review D}
  \textbf{\bibinfo{volume}{100}}, \bibinfo{pages}{043014}
  (\bibinfo{year}{2019}).

\bibitem[{\citenamefont{Blaes}(2014)}]{blaes_general_2014}
\bibinfo{author}{\bibfnamefont{O.}~\bibnamefont{Blaes}},
  \bibinfo{journal}{Space Science Reviews} \textbf{\bibinfo{volume}{183}},
  \bibinfo{pages}{21} (\bibinfo{year}{2014}), ISSN \bibinfo{issn}{1572-9672}.

\bibitem[{\citenamefont{Ryan et~al.}(2017)\citenamefont{Ryan, Gammie, Fromang,
  and Kestener}}]{ryan_resolution_2017}
\bibinfo{author}{\bibfnamefont{B.~R.} \bibnamefont{Ryan}},
  \bibinfo{author}{\bibfnamefont{C.~F.} \bibnamefont{Gammie}},
  \bibinfo{author}{\bibfnamefont{S.}~\bibnamefont{Fromang}}, \bibnamefont{and}
  \bibinfo{author}{\bibfnamefont{P.}~\bibnamefont{Kestener}},
  \bibinfo{journal}{The Astrophysical Journal} \textbf{\bibinfo{volume}{840}},
  \bibinfo{pages}{6} (\bibinfo{year}{2017}), ISSN \bibinfo{issn}{0004-637X}.

\bibitem[{\citenamefont{Shiokawa et~al.}(2012)\citenamefont{Shiokawa, Dolence,
  Gammie, and Noble}}]{shiokawa_global_2012}
\bibinfo{author}{\bibfnamefont{H.}~\bibnamefont{Shiokawa}},
  \bibinfo{author}{\bibfnamefont{J.~C.} \bibnamefont{Dolence}},
  \bibinfo{author}{\bibfnamefont{C.~F.} \bibnamefont{Gammie}},
  \bibnamefont{and} \bibinfo{author}{\bibfnamefont{S.~C.} \bibnamefont{Noble}},
  \bibinfo{journal}{The Astrophysical Journal} \textbf{\bibinfo{volume}{744}},
  \bibinfo{pages}{187} (\bibinfo{year}{2012}), ISSN \bibinfo{issn}{0004-637X}.

\bibitem[{\citenamefont{Moore et~al.}(2014)\citenamefont{Moore, Cole, and
  Berry}}]{moore_gravitational-wave_2014}
\bibinfo{author}{\bibfnamefont{C.~J.} \bibnamefont{Moore}},
  \bibinfo{author}{\bibfnamefont{R.~H.} \bibnamefont{Cole}}, \bibnamefont{and}
  \bibinfo{author}{\bibfnamefont{C.~P.~L.} \bibnamefont{Berry}},
  \bibinfo{journal}{Classical and Quantum Gravity}
  \textbf{\bibinfo{volume}{32}}, \bibinfo{pages}{015014}
  (\bibinfo{year}{2014}), ISSN \bibinfo{issn}{0264-9381}.

\bibitem[{lig()}]{ligo_and_CE_sensitivity_data}
\bibinfo{howpublished}{https://dcc.ligo.org/LIGO-P1600143/public}.

\bibitem[{\citenamefont{Yagi and Seto}(2011)}]{yagi_detector_2011}
\bibinfo{author}{\bibfnamefont{K.}~\bibnamefont{Yagi}} \bibnamefont{and}
  \bibinfo{author}{\bibfnamefont{N.}~\bibnamefont{Seto}},
  \bibinfo{journal}{Physical Review D} \textbf{\bibinfo{volume}{83}},
  \bibinfo{pages}{044011} (\bibinfo{year}{2011}).

\bibitem[{\citenamefont{Robson et~al.}(2019)\citenamefont{Robson, Cornish, and
  Liu}}]{robson_construction_2019}
\bibinfo{author}{\bibfnamefont{T.}~\bibnamefont{Robson}},
  \bibinfo{author}{\bibfnamefont{N.~J.} \bibnamefont{Cornish}},
  \bibnamefont{and} \bibinfo{author}{\bibfnamefont{C.}~\bibnamefont{Liu}},
  \bibinfo{journal}{Classical and Quantum Gravity}
  \textbf{\bibinfo{volume}{36}}, \bibinfo{pages}{105011}
  (\bibinfo{year}{2019}), ISSN \bibinfo{issn}{0264-9381}.

\bibitem[{\citenamefont{Voss and Tauris}(2003)}]{voss_galactic_2003}
\bibinfo{author}{\bibfnamefont{R.}~\bibnamefont{Voss}} \bibnamefont{and}
  \bibinfo{author}{\bibfnamefont{T.~M.} \bibnamefont{Tauris}},
  \bibinfo{journal}{Monthly Notices of the Royal Astronomical Society}
  \textbf{\bibinfo{volume}{342}}, \bibinfo{pages}{1169} (\bibinfo{year}{2003}),
  ISSN \bibinfo{issn}{0035-8711}.

\bibitem[{\citenamefont{MacFadyen and
  Woosley}(1999)}]{macfadyen_collapsars_1999}
\bibinfo{author}{\bibfnamefont{A.~I.} \bibnamefont{MacFadyen}}
  \bibnamefont{and} \bibinfo{author}{\bibfnamefont{S.~E.}
  \bibnamefont{Woosley}}, \bibinfo{journal}{The Astrophysical Journal}
  \textbf{\bibinfo{volume}{524}}, \bibinfo{pages}{262} (\bibinfo{year}{1999}).

\bibitem[{\citenamefont{MacFadyen et~al.}(2001)\citenamefont{MacFadyen,
  Woosley, and Heger}}]{macfadyen_supernovae_2001}
\bibinfo{author}{\bibfnamefont{A.~I.} \bibnamefont{MacFadyen}},
  \bibinfo{author}{\bibfnamefont{S.~E.} \bibnamefont{Woosley}},
  \bibnamefont{and} \bibinfo{author}{\bibfnamefont{A.}~\bibnamefont{Heger}},
  \bibinfo{journal}{The Astrophysical Journal} \textbf{\bibinfo{volume}{550}},
  \bibinfo{pages}{410} (\bibinfo{year}{2001}), ISSN \bibinfo{issn}{0004-637X}.

\bibitem[{\citenamefont{Heger and Woosley}(2002)}]{heger_nucleosynthetic_2002}
\bibinfo{author}{\bibfnamefont{A.}~\bibnamefont{Heger}} \bibnamefont{and}
  \bibinfo{author}{\bibfnamefont{S.~E.} \bibnamefont{Woosley}},
  \bibinfo{journal}{The Astrophysical Journal} \textbf{\bibinfo{volume}{567}},
  \bibinfo{pages}{532} (\bibinfo{year}{2002}), ISSN \bibinfo{issn}{0004-637X}.

\bibitem[{\citenamefont{Heger et~al.}(2003)\citenamefont{Heger, Fryer, Woosley,
  Langer, and Hartmann}}]{heger_how_2003}
\bibinfo{author}{\bibfnamefont{A.}~\bibnamefont{Heger}},
  \bibinfo{author}{\bibfnamefont{C.~L.} \bibnamefont{Fryer}},
  \bibinfo{author}{\bibfnamefont{S.~E.} \bibnamefont{Woosley}},
  \bibinfo{author}{\bibfnamefont{N.}~\bibnamefont{Langer}}, \bibnamefont{and}
  \bibinfo{author}{\bibfnamefont{D.~H.} \bibnamefont{Hartmann}},
  \bibinfo{journal}{The Astrophysical Journal} \textbf{\bibinfo{volume}{591}},
  \bibinfo{pages}{288} (\bibinfo{year}{2003}), ISSN \bibinfo{issn}{0004-637X}.

\bibitem[{\citenamefont{Tornatore et~al.}(2007)\citenamefont{Tornatore,
  Ferrara, and Schneider}}]{tornatore_population_2007}
\bibinfo{author}{\bibfnamefont{L.}~\bibnamefont{Tornatore}},
  \bibinfo{author}{\bibfnamefont{A.}~\bibnamefont{Ferrara}}, \bibnamefont{and}
  \bibinfo{author}{\bibfnamefont{R.}~\bibnamefont{Schneider}},
  \bibinfo{journal}{Monthly Notices of the Royal Astronomical Society}
  \textbf{\bibinfo{volume}{382}}, \bibinfo{pages}{945} (\bibinfo{year}{2007}),
  ISSN \bibinfo{issn}{0035-8711}.

\bibitem[{\citenamefont{Johnson et~al.}(2013)\citenamefont{Johnson, Dalla, and
  Khochfar}}]{johnson_first_2013}
\bibinfo{author}{\bibfnamefont{J.~L.} \bibnamefont{Johnson}},
  \bibinfo{author}{\bibfnamefont{V.~C.} \bibnamefont{Dalla}}, \bibnamefont{and}
  \bibinfo{author}{\bibfnamefont{S.}~\bibnamefont{Khochfar}},
  \bibinfo{journal}{Monthly Notices of the Royal Astronomical Society}
  \textbf{\bibinfo{volume}{428}}, \bibinfo{pages}{1857} (\bibinfo{year}{2013}),
  ISSN \bibinfo{issn}{0035-8711}.

\bibitem[{\citenamefont{Sobral et~al.}(2015)\citenamefont{Sobral, Matthee,
  Darvish, Schaerer, Mobasher, R{\"o}ttgering, Santos, and
  Hemmati}}]{sobral_evidence_2015}
\bibinfo{author}{\bibfnamefont{D.}~\bibnamefont{Sobral}},
  \bibinfo{author}{\bibfnamefont{J.}~\bibnamefont{Matthee}},
  \bibinfo{author}{\bibfnamefont{B.}~\bibnamefont{Darvish}},
  \bibinfo{author}{\bibfnamefont{D.}~\bibnamefont{Schaerer}},
  \bibinfo{author}{\bibfnamefont{B.}~\bibnamefont{Mobasher}},
  \bibinfo{author}{\bibfnamefont{H.~J.~A.} \bibnamefont{R{\"o}ttgering}},
  \bibinfo{author}{\bibfnamefont{S.}~\bibnamefont{Santos}}, \bibnamefont{and}
  \bibinfo{author}{\bibfnamefont{S.}~\bibnamefont{Hemmati}},
  \bibinfo{journal}{The Astrophysical Journal} \textbf{\bibinfo{volume}{808}},
  \bibinfo{pages}{139} (\bibinfo{year}{2015}), ISSN \bibinfo{issn}{0004-637X}.

\bibitem[{\citenamefont{Shapiro and Shibata}(2002)}]{shapiro_collapse_2002}
\bibinfo{author}{\bibfnamefont{S.~L.} \bibnamefont{Shapiro}} \bibnamefont{and}
  \bibinfo{author}{\bibfnamefont{M.}~\bibnamefont{Shibata}},
  \bibinfo{journal}{The Astrophysical Journal} \textbf{\bibinfo{volume}{577}},
  \bibinfo{pages}{904} (\bibinfo{year}{2002}), ISSN \bibinfo{issn}{0004-637X}.

\bibitem[{\citenamefont{Shapiro}(2004)}]{shapiro_collapse_2004}
\bibinfo{author}{\bibfnamefont{S.~L.} \bibnamefont{Shapiro}},
  \bibinfo{journal}{The Astrophysical Journal} \textbf{\bibinfo{volume}{610}},
  \bibinfo{pages}{913} (\bibinfo{year}{2004}), ISSN \bibinfo{issn}{0004-637X}.

\bibitem[{\citenamefont{Sun et~al.}(2017)\citenamefont{Sun, Paschalidis, Ruiz,
  and Shapiro}}]{sun_magnetorotational_2017}
\bibinfo{author}{\bibfnamefont{L.}~\bibnamefont{Sun}},
  \bibinfo{author}{\bibfnamefont{V.}~\bibnamefont{Paschalidis}},
  \bibinfo{author}{\bibfnamefont{M.}~\bibnamefont{Ruiz}}, \bibnamefont{and}
  \bibinfo{author}{\bibfnamefont{S.~L.} \bibnamefont{Shapiro}},
  \bibinfo{journal}{Physical Review D} \textbf{\bibinfo{volume}{96}},
  \bibinfo{pages}{043006} (\bibinfo{year}{2017}).

\bibitem[{\citenamefont{Uchida et~al.}(2017)\citenamefont{Uchida, Shibata,
  Yoshida, Sekiguchi, and Umeda}}]{uchida_gravitational_2017}
\bibinfo{author}{\bibfnamefont{H.}~\bibnamefont{Uchida}},
  \bibinfo{author}{\bibfnamefont{M.}~\bibnamefont{Shibata}},
  \bibinfo{author}{\bibfnamefont{T.}~\bibnamefont{Yoshida}},
  \bibinfo{author}{\bibfnamefont{Y.}~\bibnamefont{Sekiguchi}},
  \bibnamefont{and} \bibinfo{author}{\bibfnamefont{H.}~\bibnamefont{Umeda}},
  \bibinfo{journal}{Physical Review D} \textbf{\bibinfo{volume}{96}},
  \bibinfo{pages}{083016} (\bibinfo{year}{2017}).

\bibitem[{\citenamefont{Loeb and Rasio}(1994)}]{loeb_collapse_1994}
\bibinfo{author}{\bibfnamefont{A.}~\bibnamefont{Loeb}} \bibnamefont{and}
  \bibinfo{author}{\bibfnamefont{F.~A.} \bibnamefont{Rasio}},
  \bibinfo{journal}{The Astrophysical Journal} \textbf{\bibinfo{volume}{432}},
  \bibinfo{pages}{52} (\bibinfo{year}{1994}).

\bibitem[{\citenamefont{Shapiro}(2003)}]{shapiro_relativistic_2003}
\bibinfo{author}{\bibfnamefont{S.~L.} \bibnamefont{Shapiro}},
  \bibinfo{journal}{AIP Conference Proceedings} \textbf{\bibinfo{volume}{686}},
  \bibinfo{pages}{50} (\bibinfo{year}{2003}), ISSN \bibinfo{issn}{0094-243X}.

\bibitem[{\citenamefont{Koushiappas et~al.}(2004)\citenamefont{Koushiappas,
  Bullock, and Dekel}}]{koushiappas_massive_2004}
\bibinfo{author}{\bibfnamefont{S.~M.} \bibnamefont{Koushiappas}},
  \bibinfo{author}{\bibfnamefont{J.~S.} \bibnamefont{Bullock}},
  \bibnamefont{and} \bibinfo{author}{\bibfnamefont{A.}~\bibnamefont{Dekel}},
  \bibinfo{journal}{Monthly Notices of the Royal Astronomical Society}
  \textbf{\bibinfo{volume}{354}}, \bibinfo{pages}{292} (\bibinfo{year}{2004}),
  ISSN \bibinfo{issn}{0035-8711}.

\bibitem[{\citenamefont{Shapiro}(2005)}]{shapiro_spin_2005}
\bibinfo{author}{\bibfnamefont{S.~L.} \bibnamefont{Shapiro}},
  \bibinfo{journal}{The Astrophysical Journal} \textbf{\bibinfo{volume}{620}},
  \bibinfo{pages}{59} (\bibinfo{year}{2005}), ISSN \bibinfo{issn}{0004-637X}.

\bibitem[{\citenamefont{Begelman et~al.}(2006)\citenamefont{Begelman,
  Volonteri, and Rees}}]{begelman_formation_2006}
\bibinfo{author}{\bibfnamefont{M.~C.} \bibnamefont{Begelman}},
  \bibinfo{author}{\bibfnamefont{M.}~\bibnamefont{Volonteri}},
  \bibnamefont{and} \bibinfo{author}{\bibfnamefont{M.~J.} \bibnamefont{Rees}},
  \bibinfo{journal}{Monthly Notices of the Royal Astronomical Society}
  \textbf{\bibinfo{volume}{370}}, \bibinfo{pages}{289} (\bibinfo{year}{2006}),
  ISSN \bibinfo{issn}{0035-8711}.

\bibitem[{\citenamefont{Lodato and Natarajan}(2006)}]{lodato_supermassive_2006}
\bibinfo{author}{\bibfnamefont{G.}~\bibnamefont{Lodato}} \bibnamefont{and}
  \bibinfo{author}{\bibfnamefont{P.}~\bibnamefont{Natarajan}},
  \bibinfo{journal}{Monthly Notices of the Royal Astronomical Society}
  \textbf{\bibinfo{volume}{371}}, \bibinfo{pages}{1813} (\bibinfo{year}{2006}),
  ISSN \bibinfo{issn}{0035-8711}.

\bibitem[{\citenamefont{Begelman}(2010)}]{begelman_evolution_2010}
\bibinfo{author}{\bibfnamefont{M.~C.} \bibnamefont{Begelman}},
  \bibinfo{journal}{Monthly Notices of the Royal Astronomical Society}
  \textbf{\bibinfo{volume}{402}}, \bibinfo{pages}{673} (\bibinfo{year}{2010}),
  ISSN \bibinfo{issn}{0035-8711}.

\bibitem[{\citenamefont{Haiman}(2013)}]{haiman_formation_2013}
\bibinfo{author}{\bibfnamefont{Z.}~\bibnamefont{Haiman}},
  \textbf{\bibinfo{volume}{396}}, \bibinfo{pages}{293} (\bibinfo{year}{2013}).

\bibitem[{\citenamefont{Latif and Ferrara}(2016/ed)}]{latif_formation_2016}
\bibinfo{author}{\bibfnamefont{M.~A.} \bibnamefont{Latif}} \bibnamefont{and}
  \bibinfo{author}{\bibfnamefont{A.}~\bibnamefont{Ferrara}},
  \bibinfo{journal}{Publications of the Astronomical Society of Australia}
  \textbf{\bibinfo{volume}{33}} (\bibinfo{year}{2016/ed}), ISSN
  \bibinfo{issn}{1323-3580, 1448-6083}.

\bibitem[{\citenamefont{Smith et~al.}(2017)\citenamefont{Smith, Bromm, and
  Loeb}}]{smith_first_2017}
\bibinfo{author}{\bibfnamefont{A.}~\bibnamefont{Smith}},
  \bibinfo{author}{\bibfnamefont{V.}~\bibnamefont{Bromm}}, \bibnamefont{and}
  \bibinfo{author}{\bibfnamefont{A.}~\bibnamefont{Loeb}},
  \bibinfo{journal}{Astronomy \& Geophysics} \textbf{\bibinfo{volume}{58}},
  \bibinfo{pages}{3.22} (\bibinfo{year}{2017}), ISSN \bibinfo{issn}{1366-8781}.

\end{thebibliography}

\end{document}